\documentclass[aps,prd,preprint,nofootinbib,11pt]{revtex4}
\usepackage{amsfonts}
\usepackage{mathrsfs}
\usepackage{graphicx}
\usepackage{float}
\usepackage[T1]{fontenc}
\usepackage{booktabs,multirow,array}
\usepackage{amsmath}
\usepackage{amssymb}
\usepackage{multirow}
\usepackage{subfigure}
\usepackage{epsfig}
\usepackage{graphicx}
\usepackage{color}
\newcommand{\bqa}{\begin{eqnarray}}
\newcommand{\eqa}{\end{eqnarray}}
\newcommand{\beq}{\begin{equation}}
\newcommand{\eeq}{\end{equation}}
\allowdisplaybreaks[1]
\graphicspath{{fig/}{dia/}} \DeclareGraphicsExtensions{.eps}

\begin{document}

\title{QCD sum rule predictions on gluonic tetraquark states with $J^{PC}=0^{+-},0^{--}$ and $1^{\pm \pm}$}
\author{Chun-Meng Tang$^{1}$}
\author{Chun-Gui Duan$^{2,3}$}
\email{duancg@hebtu.edu.cn}
\author{Liang Tang$^{1}$}
\email{tangl@hebtu.edu.cn}
\author{Cong-Feng Qiao$^{4}$}
\email{qiaocf@ucas.ac.cn}
\affiliation{$^1$ College of Physics and Hebei Key Laboratory of Photophysics Research and Application,
Hebei Normal University, Shijiazhuang 050024, China \\
$^2$College of Physics and Hebei Advanced Thin Film Laboratory, Hebei Normal University, Shijiazhuang 050024, China\\
$^3$ Hebei Key Laboratory of Physics and Energy Technology, North China Electric Power University, Baoding 071000, China\\
$^4$ School of Physical Sciences, University of Chinese Academy of Sciences,
Yuquan Road 19A, Beijing 100049, China
}

%\author{~\\~\\}

%\affiliation{}

\begin{abstract}
\vspace{0.3cm}
In this work, we present a systematic calculation of the mass spectrum for tetraquark hybrid states, focusing on the $8_{[c\bar{c}]}\otimes 8_{[G]}\otimes 8_{[c\bar{c}]}$ color configuration, within the framework of QCD sum rules. As an extension of our previous work on $0^{++}$ and $0^{-+}$ states, we now construct 18 distinct interpolating currents with $J^{PC} = 0^{+-}$, $0^{--}$, and $1^{\pm\pm}$. Using operator product expansion (OPE) techniques and including nonperturbative contributions up to dimension six, we obtain key results: for the $0^{+-}$, $1^{--}$, and $1^{-+}$ states, the predicted masses lie in the range of $7.2-7.3$ GeV, while the $1^{+-}$ and $1^{++}$ states have slightly lower masses, between 6.9 and 7.1 GeV. These predictions provide strong support for the possible existence of an $8_{[c\bar{c}]}\otimes 8_{[G]}\otimes 8_{[c\bar{c}]}$ component within the di-$J/\psi$ structure reported by LHCb. Moreover, our analogous calculations for tetrabottom hybrid states yield mass ranges of $19.4-19.5$ GeV (for $0^{+-}$, $1^{--}$, and $1^{-+}$) and $19.2-19.3$ GeV (for $1^{+-}$ and $1^{++}$), offering crucial references for future searches.
\end{abstract}
\maketitle
\newpage

\section{Introduction}
The field of hadronic physics is currently in a pivotal stage of development. Since Gell-Mann and Zweig first postulated the existence of multiquark states in the formulation of the quark model~\cite{Gell-Mann:1964ewy,Zweig:1964ruk}, theoretical physicists have continuously explored exotic hadronic structures beyond conventional mesons ($q\bar{q}$) and baryons ($qqq$). Quantum chromodynamics (QCD) provides a rigorous theoretical framework for such unconventional hadronic states, including multiquark states, hadronic molecules, hybrid states, and glueballs~\cite{Gross:1973id,Politzer:1973fx,Wilson:1974sk}.

With the advent of the 21st century, significant progress in high energy experimental techniques has led to the discovery of a plethora of charmoniumlike exotic states (known as XYZ states) and $P_c$ states~\cite{Belle:2003nnu, BaBar:2005hhc, Belle:2011aa, BESIII:2013ris, Belle:2013yex, LHCb:2015yax, LHCb:2019kea}. To date, more than 60 new hadronic candidates have been reported, reflecting the continued expansion of the hadronic spectrum. Many of these newly observed resonances cannot be explained within the conventional quark model, motivating theorists to develop innovative structural paradigms. A major ongoing challenge is distinguishing between loosely bound hadronic molecules and compact multiquark states, especially in studies of XYZ and $P_c$ states, where the presence of light quarks adds complexity to both experimental and theoretical analyses.

At the same time, substantial progress has been made in the search for hybrid states\textemdash exotic hadrons that consist of quark-antiquark pairs coupled to dynamical gluonic fields. These states have become high-priority targets in large-scale experiments such as BESIII and LHCb. Particularly notable are the recent observations by the BESIII collaboration: the $1^{-+}$ exotic state $\eta_{1}(1855)$, identified in the $\eta\eta'$ invariant mass spectrum with a significance of 19$\sigma$~\cite{BESIII:2022riz, BESIII:2022iwi}, as well as earlier candidates with exotic quantum numbers, $\pi_{1}(1400)$~\cite{IHEP-Brussels-LosAlamos-AnnecyLAPP:1988iqi} and $\pi_{1}(1600)$~\cite{E852:2001ikk, COMPASS:2009xrl}. These findings provide strong experimental support for the essential role of gluonic degrees of freedom in hadronic structure. Such landmark discoveries not only drive hadronic physics toward more profound investigations but also hold the potential to greatly enhance our understanding of nonperturbative QCD.

Recent years have witnessed landmark breakthroughs in the experimental investigation of fully charm tetraquark states. In 2020, the LHCb collaboration first reported a broad structure in the 6.2-6.8 GeV range and a narrow resonance $X(6900)$ with a significance exceeding 5$\sigma$ in the $J/\psi J/\psi$ invariant mass spectrum using proton-proton collision data at $\sqrt{s}$=7, 8, and 13 TeV, while suggesting possible evidence for new structures near 7.2 GeV~\cite{LHCb:2020bwg}. This observation was subsequently corroborated through independent confirmations: the ATLAS collaboration not only confirmed the $X(6900)$ with higher significance but also identified new structures, including the $X(6400)$ and $X(6600)$~\cite{ATLAS:2023bft}; the CMS collaboration observed the $X(6900)$ signal with a significance of 9.4$\sigma$ and further resolved two additional resonances, the $X(6600)$ and $X(7200)$~\cite{CMS:2023owd}. Particularly noteworthy is that ATLAS observed signals of the $X(6900)$ and $X(7200)$ at a significance level of 4.7$\sigma$ through the analysis of the $J/\psi \psi(2S)$ decay channel~\cite{ATLAS:2023bft}, thereby opening a new decay channel for probing tetraquark states. These pivotal findings not only substantiate the existence of fully charm tetraquarks but also reveal multiple potential excited states within the 6.2-7.2 GeV mass range, providing crucial experimental constraints for theoretical studies. Table \ref{tab1} systematically compiles the latest observational results from these experiments.

\begin{table}[htb]
\centering
\caption{Experimental observations of fully charm tetraquark state candidates.}
\label{tab:tetraquark}
\begin{tabular}{llccc}
\toprule
Collaborations & Resonances & Masses (MeV) & Widths (MeV) & Observed Channels \\
\midrule
LHCb (model I) & \multirow{2}{*}{$X(6900)$} & 6905 $\pm$ 11 $\pm$ 7 & 80 $\pm$ 19 $\pm$ 33 & \multirow{2}{*}{di--J/$\psi$ \cite{LHCb:2020bwg}} \\
LHCb (model II) & & 6886 $\pm$ 11 $\pm$ 1 & 168 $\pm$ 33 $\pm$ 69 & \\
\hline
\multirow{3}{*}{ATLAS (model A)} & $X(6400)$ & 6410 $\pm$ 80$^{+60}_{-30}$ & 590 $\pm$ 350$^{+120}_{-200}$ & \multirow{3}{*}{di--J/$\psi$ \cite{ATLAS:2023bft}} \\
 & $X(6600)$ & 6630 $\pm$ 50$^{+80}_{-10}$ & 350 $\pm$ 110$^{+110}_{-40}$ & \\
 & $X(6900)$ & 6860 $\pm$ 30$^{+10}_{-20}$ & 110 $\pm$ 50$^{+20}_{-10}$ & \\
\hline
\multirow{2}{*}{ATLAS (model B)} & $X(6600)$ & 6650 $\pm$ 20$^{+30}_{-20}$ & 440 $\pm$ 50$^{+60}_{-50}$ & \multirow{2}{*}{di--J/$\psi$ \cite{ATLAS:2023bft}} \\
 & $X(6900)$ & 6910 $\pm$ 10 $\pm$ 10 & 150 $\pm$ 30 $\pm$ 10 & \\
\hline
ATLAS (model $\beta$) & $X(6900)$ & 6960 $\pm$ 50 $\pm$ 30 & 510 $\pm$ 170$^{+110}_{-100}$ & \multirow{2}{*}{J/$\psi\psi$(2S) \cite{ATLAS:2023bft}} \\
ATLAS (model $\alpha$) & $X(7200)$ & 7220 $\pm$ 30$^{+10}_{-40}$ & 90 $\pm$ 60$^{+60}_{-50}$ & \\
\hline
\multirow{3}{*}{CMS (No-interference)} & $X(6600)$ & 6552 $\pm$ 10 $\pm$ 12 & 124$^{+32}_{-32}$ $\pm$ 33 & \multirow{3}{*}{di--J/$\psi$ \cite{CMS:2023owd}} \\
 & $X(6900)$ & 6927 $\pm$ 9 $\pm$ 4 & 122$^{+24}_{-21}$ $\pm$ 18 & \\
 & $X(7200)$ & 7287$^{+20}_{-18}$ $\pm$ 5 & 95$^{+59}_{-40}$ $\pm$ 19 & \\
\hline
\multirow{3}{*}{CMS (Interference)} & $X(6600)$ & 6638$^{+43+16}_{-38-31}$ & 440$^{+230+110}_{-200-240}$ & \multirow{3}{*}{di--J/$\psi$ \cite{CMS:2023owd}} \\
 & $X(6900)$ & 6847$^{+44+48}_{-28-20}$ & 191$^{+66+25}_{-17}$ & \\
 & $X(7200)$ & 7134$^{+48+41}_{-25-15}$ & 97$^{+40+29}_{-29-26}$ & \\
\bottomrule
\end{tabular}
\label{tab1}
\end{table}

The investigation of fully charm tetraquark states has emerged as a vanguard pursuit in high energy physics, with their distinctive strong interaction mechanisms furnishing a crucial portal into the nonperturbative regime of quantum chromodynamics (QCD). Theoretical physicists have conducted systematic investigations on this particular hadronic state within diverse theoretical frameworks~\cite{Anwar:2017toa, Karliner:2016zzc, Debastiani:2017msn, Wu:2016vtq, Liu:2019zuc, Yu:2022lak, Esposito:2018cwh, Bai:2016int, Dong:2022sef, Yan:2023lvm, Wu:2024tif, Bedolla:2019zwg, Jin:2020jfc, Liu:2021rtn,Mutuk:2021hmi,Faustov:2020qfm,Lu:2020cns,Karliner:2020dta, Giron:2020wpx,Chao:2020dml,Maiani:2020pur,Richard:2020hdw,Zhu:2020xni,Li:2021ygk, Ke:2021iyh,Becchi:2020uvq,Zhang:2020xtb,Wang:2020dlo,Chen:2020xwe,Zhao:2020nwy, Wang:2020ols,Wang:2021kfv,Gordillo:2020sgc,liu:2020eha,Sonnenschein:2020nwn, Zhao:2020jvl,Deng:2020iqw,Chen:2022sbf,Asadi:2021ids,Zhang:2022qtp,Wang:2022xja, Albuquerque:2020hio,Wang:2021mma,Wan:2020fsk,Yang:2020wkh,Tang:2024zvf,Lloyd:2003yc, Barnea:2006sd,Wang:2019rdo,Heupel:2012ua,Weng:2020jao,Berezhnoy:2011xn,Berezhnoy:2011xy, Feng:2020riv,Zhang:2020hoh,Wang:2020wrp,Wang:2022jmb,Wang:2020tpt,Maciula:2020wri, Zhu:2020snb,Eichmann:2020oqt,Gong:2020bmg,Dong:2020nwy,Chen:2016jxd,Wang:2017jtz, Chen:2018cqz,Wang:2018poa,Tang:2024kmh,Agaev:2023wua,Agaev:2023gaq,Agaev:2023ruu, Agaev:2023rpj,Agaev:2023ara,Dong:2021lkh,Gong:2022hgd,Wang:2021wjd,Liu:2020tqy, Yang:2021hrb,Chiu:2005ey,Santowsky:2021bhy,Liang:2021fzr,Guo:2020pvt,Cao:2020gul, Nefediev:2021pww,Zhuang:2021pci}. Conventional approaches including potential quark models~\cite{Anwar:2017toa,Karliner:2016zzc,Debastiani:2017msn,Wu:2016vtq,Liu:2019zuc,Yu:2022lak, Esposito:2018cwh,Bai:2016int,Dong:2022sef,Yan:2023lvm,Wu:2024tif,Bedolla:2019zwg, Jin:2020jfc,Liu:2021rtn,Mutuk:2021hmi}, QCD sum rules~\cite{Zhang:2020xtb, Wang:2020dlo, Wang:2020ols, Wang:2022xja, Albuquerque:2020hio,Wang:2021mma,Wan:2020fsk,Yang:2020wkh,Tang:2024zvf, Chen:2016jxd, Wang:2017jtz, Wang:2018poa, Tang:2024kmh, Agaev:2023wua, Agaev:2023gaq} and lattice QCD calculations~\cite{Chiu:2005ey} have elucidated fundamental spectroscopic characteristics, while dynamical rescattering mechanisms~\cite{Wang:2020wrp,Wang:2022jmb,Wang:2020tpt}, Bethe-Salpeter equations~\cite{Zhu:2020xni,Li:2021ygk, Ke:2021iyh,Heupel:2012ua,Santowsky:2021bhy}, and coupled-channel final state interaction theories~\cite{Gong:2020bmg, Dong:2020nwy, Dong:2021lkh, Gong:2022hgd, Liang:2021fzr,Guo:2020pvt,Nefediev:2021pww,Cao:2020gul} have further deepened comprehension of bound-state formation dynamics. Nevertheless, extant theoretical frameworks exhibit marked discordance when confronting experimentally observed resonance spectra and decay properties, a predicament underscoring the intricate internal architecture of  fully charm tetraquarks. Researchers have posited multiple structural hypotheses encompassing compact tetraquark configurations~\cite{Bedolla:2019zwg, Jin:2020jfc,Liu:2021rtn,Mutuk:2021hmi,Faustov:2020qfm,Lu:2020cns,Karliner:2020dta, Giron:2020wpx,Chao:2020dml,Maiani:2020pur,Richard:2020hdw,Zhu:2020xni,Li:2021ygk, Ke:2021iyh,Becchi:2020uvq,Zhang:2020xtb,Wang:2020dlo,Chen:2020xwe,Zhao:2020nwy, Wang:2020ols,Wang:2021kfv,Gordillo:2020sgc,liu:2020eha,Sonnenschein:2020nwn, Zhao:2020jvl,Deng:2020iqw,Chen:2022sbf,Asadi:2021ids,Zhang:2022qtp,Wang:2022xja}, and tetraquark-molecule mixing structures~\cite{Santowsky:2021bhy}. More innovative proposals include hidden-color octet configurations~\cite{Wang:2021mma,Yang:2020wkh}, dynamically generated resonance poles~\cite{Gong:2020bmg,Gong:2022hgd,Liang:2021fzr,Guo:2020pvt}, cusp kinematic effects~\cite{Zhuang:2021pci}, Higgs-like boson analogies~\cite{Zhu:2020snb}. The introduction of unconventional interpretations---particularly the gluonic tetracharm configurations~\cite{Wan:2020fsk,Tang:2024kmh}---has not only enriched the theoretical dimensions but also yielded specific predictions that await experimental confirmation.

Building upon our prior investigation into tetracharm hybrid states (quantum numbers \(0^{++}\) and \(0^{-+}\)), this study delves deeper into this intriguing domain. In our previous work, we pioneered the construction of a novel hadronic configuration featuring color-octet charm quark pairs coupled with a color-octet gluon (\(8_{[c\bar{c}]} \otimes 8_{[G]} \otimes 8_{[c\bar{c}]}\))~\cite{Tang:2024kmh}, with theoretical calculations suggesting that the \(X(6900)\) and \(X(7200)\) resonances observed by LHCb might correspond to the \(0^{++}\) and \(0^{-+}\) states of this configuration, respectively. To refine this theoretical framework, the current work systematically extends the scope to encompass all possible quantum number states with spins 0 and 1. Our research applies the QCD sum rules (QCDSR) approach. Extensive research has demonstrated that this method has yielded a series of pivotal breakthroughs in hadronic spectroscopy~\cite{Yang:2020wkh, Tang:2024zvf, Qiao:2010zh, Chen:2013zia, Chen:2013eha, Wang:2024hvp, Shifman:1978bx, Albuquerque:2013ija, Wang:2013vex, Govaerts:1984hc, Reinders:1984sr, P.Col, Narison:1989aq, Tang:2021zti, Qiao:2014vva, Qiao:2015iea, Tang:2019nwv, Wan:2019ake, Wan:2020oxt, Li:2024ctd, Wan:2021vny, Zhao:2023imq, Wan:2022xkx, Zhang:2022obn, Wan:2022uie, Yin:2021cbb, Wan:2023epq, Wan:2024dmi, Wan:2024cpc}. In the practical implementation of QCDSR, the initial step involves constructing appropriate current operators based on the quantum characteristics of the studied system. Using these current operators, we define the two-point correlation function for studying hadronic properties. Notably, these correlation functions possess dual representations: one expressed in terms of QCD's fundamental degrees of freedom (quarks and gluons), and the other characterized by hadronic physical observables. Through a carefully designed mathematical matching procedure that equates these two representations, robust sum rules can be established, allowing for the precise extraction of fundamental hadronic mass parameters.

This paper is organized as follows. Section II outlines the theoretical framework and foundational concepts. Section III details the computational methodology and presents the corresponding numerical results. The decay processes are analyzed in Section IV, and the study concludes with a general discussion.

\section{Formalism}
This section presents the theoretical foundations of Quantum Chromodynamics Sum Rules (QCDSR), an important methodology in modern hadronic physics. The starting point of our derivation is the two-point correlation function. This function is formulated in terms of local composite operators with definite Lorentz structures, including the scalar current $j(x)$ and the vector current $j_\mu(x)$. These correlation functions form the theoretical foundation for the subsequent operator product expansion and spectral density analysis, given by
\begin{eqnarray}
\Pi(q)&=&i\int d^{4}xe^{iq\cdot x}\langle 0|T\{j(x),j^{\dagger}(0)\}|0 \rangle,\label{Pi}\\
\Pi_{\mu\nu}(q)&=&i\int d^{4}xe^{iq\cdot x}\langle 0|T\{j_{\mu}(x),j_{\nu}^{\dagger}(0)\}|0 \rangle,
\label{Pimunu}
\end{eqnarray}
where, $j(x)$ and $j_{\mu}(x)$ represent the hadronic current operators for $J=0$ and $J=1$ states, respectively.

The interpolating current operators investigated in this work consist of two color-octet $Q\bar{Q}$ components coupled with a gluon field. For the $J^{PC}=0^{--}$ and $0^{+-}$ tetraquark hybrid states, the current can be constructed in the following form
\begin{eqnarray}
j^{0^{--}}(x) &=& g_{s}f^{abc}\left[ \bar{Q}^{j}(x)\gamma^{\mu}\gamma_{5}(t^{a})_{jk}Q^{k}(x) \right]G^{b}_{\mu\nu}(x) \left[ \bar{Q}^{m}(x)\gamma^{\nu}(t^{c})_{mn}Q^{n}(x) \right]\; \label{current-1} ,\\
j^{0^{+-}}(x) &=& g_{s}f^{abc}\left[ \bar{Q}^{j}(x)\gamma^{\mu}\gamma_{5}(t^{a})_{jk}Q^{k}(x) \right]\tilde{G}^{b}_{\mu\nu}(x) \left[ \bar{Q}^{m}(x)\gamma^{\nu}\gamma_{5}(t^{c})_{mn}Q^{n}(x) \right]\; \label{current-2} .
\end{eqnarray}
It should be noted that the tetraquark hybrid states with quantum numbers $0^{++}$ and $0^{-+}$ have been investigated in our previous work~\cite{Tang:2024kmh}, which serves as the theoretical foundation for the present study.

For hadronic states with quantum numbers $1^{--}$, the corresponding currents assume the following representation
\begin{eqnarray}
j^{1^{--}}_{A}(x) &=& g_{s}f^{abc}\left[ \bar{Q}^{j}(x)\gamma^{\nu}(t^{a})_{jk}Q^{k}(x) \right]G^{b}_{\mu\nu}(x) \left[ \bar{Q}^{m}(x)(t^{c})_{mn}Q^{n}(x) \right]\; \label{current-3} ,\\
j^{1^{--}}_{B}(x) &=& g_{s}f^{abc}\left[ \bar{Q}^{j}(x)\gamma^{\nu}(t^{a})_{jk}Q^{k}(x) \right]\tilde{G}^{b}_{\mu\nu}(x) \left[ \bar{Q}^{m}(x)\gamma_{5}(t^{c})_{mn}Q^{n}(x) \right]\; \label{current-4} ,\\
j^{1^{--}}_{C}(x) &=& g_{s}d^{abc}\left[ \bar{Q}^{j}(x)\gamma^{\nu}\gamma_{5}(t^{a})_{jk}Q^{k}(x) \right]\tilde{G}^{b}_{\mu\nu}(x) \left[ \bar{Q}^{m}(x)(t^{c})_{mn}Q^{n}(x) \right]\; \label{current-5} ,\\
j^{1^{--}}_{D}(x) &=& g_{s}d^{abc}\left[ \bar{Q}^{j}(x)\gamma^{\nu}\gamma_{5}(t^{a})_{jk}Q^{k}(x) \right]G^{b}_{\mu\nu}(x) \left[ \bar{Q}^{m}(x)\gamma_{5}(t^{c})_{mn}Q^{n}(x) \right]\; \label{current-6} .
\end{eqnarray}

For hadronic states with quantum numbers \(1^{+-}\), the corresponding currents are formulated as follows
\begin{eqnarray}
j^{1^{+-}}_{A}(x) &=& g_{s}f^{abc}\left[ \bar{Q}^{j}(x)\gamma^{\nu}(t^{a})_{jk}Q^{k}(x) \right]\tilde{G}^{b}_{\mu\nu}(x) \left[ \bar{Q}^{m}(x)(t^{c})_{mn}Q^{n}(x) \right]\; \label{current-7} ,\\
j^{1^{+-}}_{B}(x) &=& g_{s}f^{abc}\left[ \bar{Q}^{j}(x)\gamma^{\nu}(t^{a})_{jk}Q^{k}(x) \right]G^{b}_{\mu\nu}(x) \left[ \bar{Q}^{m}(x)\gamma_{5}(t^{c})_{mn}Q^{n}(x) \right]\; \label{current-8} ,\\
j^{1^{+-}}_{C}(x) &=& g_{s}d^{abc}\left[ \bar{Q}^{j}(x)\gamma^{\nu}\gamma_{5}(t^{a})_{jk}Q^{k}(x) \right]G^{b}_{\mu\nu}(x) \left[ \bar{Q}^{m}(x)(t^{c})_{mn}Q^{n}(x) \right]\; \label{current-9} ,\\
j^{1^{+-}}_{D}(x) &=& g_{s}d^{abc}\left[ \bar{Q}^{j}(x)\gamma^{\nu}\gamma_{5}(t^{a})_{jk}Q^{k}(x) \right]\tilde{G}^{b}_{\mu\nu}(x) \left[ \bar{Q}^{m}(x)\gamma_{5}(t^{c})_{mn}Q^{n}(x) \right]\; \label{current-10} .
\end{eqnarray}

For hadronic states bearing exotic quantum numbers \(1^{-+}\), the corresponding interpolating currents are explicitly constructed as follows
\begin{eqnarray}
j^{1^{-+}}_{A}(x) &=& g_{s}d^{abc}\left[ \bar{Q}^{j}(x)\gamma^{\nu}(t^{a})_{jk}Q^{k}(x) \right]G^{b}_{\mu\nu}(x) \left[ \bar{Q}^{m}(x)(t^{c})_{mn}Q^{n}(x) \right]\; \label{current-11} ,\\
j^{1^{-+}}_{B}(x) &=& g_{s}d^{abc}\left[ \bar{Q}^{j}(x)\gamma^{\nu}(t^{a})_{jk}Q^{k}(x) \right]\tilde{G}^{b}_{\mu\nu}(x) \left[ \bar{Q}^{m}(x)\gamma_{5}(t^{c})_{mn}Q^{n}(x) \right]\; \label{current-12} ,\\
j^{1^{-+}}_{C}(x) &=& g_{s}f^{abc}\left[ \bar{Q}^{j}(x)\gamma^{\nu}\gamma_{5}(t^{a})_{jk}Q^{k}(x) \right]\tilde{G}^{b}_{\mu\nu}(x) \left[ \bar{Q}^{m}(x)(t^{c})_{mn}Q^{n}(x) \right]\; \label{current-13} ,\\
j^{1^{-+}}_{D}(x) &=& g_{s}f^{abc}\left[ \bar{Q}^{j}(x)\gamma^{\nu}\gamma_{5}(t^{a})_{jk}Q^{k}(x) \right]G^{b}_{\mu\nu}(x) \left[ \bar{Q}^{m}(x)\gamma_{5}(t^{c})_{mn}Q^{n}(x) \right]\; \label{current-14} .
\end{eqnarray}

For hadronic states with quantum numbers \(1^{++}\), the associated interpolating currents are rigorously constructed as follows
\begin{eqnarray}
j^{1^{++}}_{A}(x) &=& g_{s}d^{abc}\left[ \bar{Q}^{j}(x)\gamma^{\nu}(t^{a})_{jk}Q^{k}(x) \right]\tilde{G}^{b}_{\mu\nu}(x) \left[ \bar{Q}^{m}(x)(t^{c})_{mn}Q^{n}(x) \right]\; \label{current-15} ,\\
j^{1^{++}}_{B}(x) &=& g_{s}d^{abc}\left[ \bar{Q}^{j}(x)\gamma^{\nu}(t^{a})_{jk}Q^{k}(x) \right]G^{b}_{\mu\nu}(x) \left[ \bar{Q}^{m}(x)\gamma_{5}(t^{c})_{mn}Q^{n}(x) \right]\; \label{current-16} ,\\
j^{1^{++}}_{C}(x) &=& g_{s}f^{abc}\left[ \bar{Q}^{j}(x)\gamma^{\nu}\gamma_{5}(t^{a})_{jk}Q^{k}(x) \right]G^{b}_{\mu\nu}(x) \left[ \bar{Q}^{m}(x)(t^{c})_{mn}Q^{n}(x) \right]\; \label{current-17} ,\\
j^{1^{++}}_{D}(x) &=& g_{s}f^{abc}\left[ \bar{Q}^{j}(x)\gamma^{\nu}\gamma_{5}(t^{a})_{jk}Q^{k}(x) \right]\tilde{G}^{b}_{\mu\nu}(x) \left[ \bar{Q}^{m}(x)\gamma_{5}(t^{c})_{mn}Q^{n}(x) \right]\; \label{current-18} ,
\end{eqnarray}
here, $g_{s}$ denotes the strong coupling constant, $Q$ represents either a charm or bottom quark, while $f^{abc}$ and $d^{abc}$ correspond to the totally antisymmetric and symmetric structure constants of the SU(3) group, respectively. The indices $\mu$ and $\nu$ denote Lorentz components, while $j$, $k$, $m$, $n$ (values 1, 2, 3) and $a$, $b$, $c$ (values 1 to 8) represent color indices in the fundamental and adjoint representations of SU(3), respectively. The generators $t^{a}$ are defined by the Gell-Mann matrices as $t^{a} = \lambda^{a}/2$. The gluon field strength tensor is denoted by $G^{b}_{\mu\nu}$, and its dual is defined as $\tilde{G}^{b}_{\mu\nu} = \frac{1}{2}\epsilon_{\mu\nu\alpha\beta}G^{b,\alpha\beta}$.

Building upon the current operators delineated in Eqs.\eqref{current-1}-\eqref{current-18}, the two-point correlation functions specified in Eqs.\eqref{Pi} and \eqref{Pimunu} can be computed separately through the operator product expansion (OPE) approach and the phenomenological representation. On the OPE side, these correlation functions are expressible via dispersion relations as follows
\begin{eqnarray}\label{Pi-OPE}
  \Pi^{\text{OPE}}(q^2) = \int_{(4m_{Q})^2}^\infty ds \frac{\rho^{\text{OPE}}(s)}{s - q^2},
\end{eqnarray}
herein, $\rho^{\text{OPE}}(s) = \text{Im} [\Pi^{\text{OPE}}(s)]/\pi$ designates the spectral density on the OPE side. When incorporating condensate terms up to dimension-six, the spectral density $\rho^{\text{OPE}}$ admits a systematic expansion expressed as
\begin{eqnarray}\label{rho-OPE}
  \rho^{\text{OPE}}(s) &=& \rho^{\text{pert}}(s) + \rho^{\langle GG \rangle}(s)+ \rho^{\langle GGG \rangle}(s).
\end{eqnarray}

To compute the spectral density on the OPE side (Eq. \eqref{rho-OPE}), we employ the full propagator of heavy quarks $S_{jk}^{Q}(p)$
\begin{eqnarray}
S^Q_{j k}(p) \! \! & = & \! \! \frac{i \delta_{j k}(p\!\!\!\slash + m_Q)}{p^2 - m_Q^2} - \frac{i}{4}g_{s} \frac{t^a_{j k} G^a_{\alpha\beta} }{(p^2 - m_Q^2)^2} [\sigma^{\alpha \beta}
(p\!\!\!\slash + m_Q)
+ (p\!\!\!\slash + m_Q) \sigma^{\alpha \beta}] \nonumber \\ &-&
\frac{i}{4}g_{s}^{2}(t^{a}t^{b})_{jk} G^{a}_{\alpha\beta}G^{b}_{\mu\nu}\frac{(p\!\!\!\slash + m_Q)}{(p^2 - m_Q^2)^5}(f^{\alpha\beta\mu\nu}+f^{\alpha\mu\beta\nu}+f^{\alpha\mu\nu\beta}) (p\!\!\!\slash + m_Q)\nonumber \\ &+& \frac{i \delta_{j k}}{48} \bigg\{ \frac{(p\!\!\!\slash +
m_Q) [p\!\!\!\slash (p^2 - 3 m_Q^2) + 2 m_Q (2 p^2 - m_Q^2)] }{(p^2 - m_Q^2)^6}
\times (p\!\!\!\slash + m_Q)\bigg\} \langle g_s^3 G^3 \rangle \; ,
\end{eqnarray}
where the subscripts $j$ and $k$ denote the color indices of heavy quarks. Herein, we define the tensor structure $f^{\alpha\beta\mu\nu}\equiv \gamma^{\alpha}(p\!\!\!\!\slash + m_Q)\gamma^{\beta}(p\!\!\!\!\slash + m_Q)\gamma^{\mu}(p\!\!\!\!\slash + m_Q)\gamma^{\nu}$. For comprehensive details regarding the aforementioned propagator, readers may consult references~\cite{Wang:2013vex, Reinders:1984sr, Albuquerque:2012jbz}.

By applying the Borel transformation to Eq.~\eqref{Pi-OPE}, we derive
\begin{eqnarray}\label{Pi-MB}
  \Pi^{\text{OPE}}(M_B^2) = \int_{(4m_{Q})^2}^\infty ds \rho^{\text{OPE}}(s) e^{-s/M_B^2}.
\end{eqnarray}

\begin{figure}[ht]
  \centering
  \includegraphics[width=12cm]{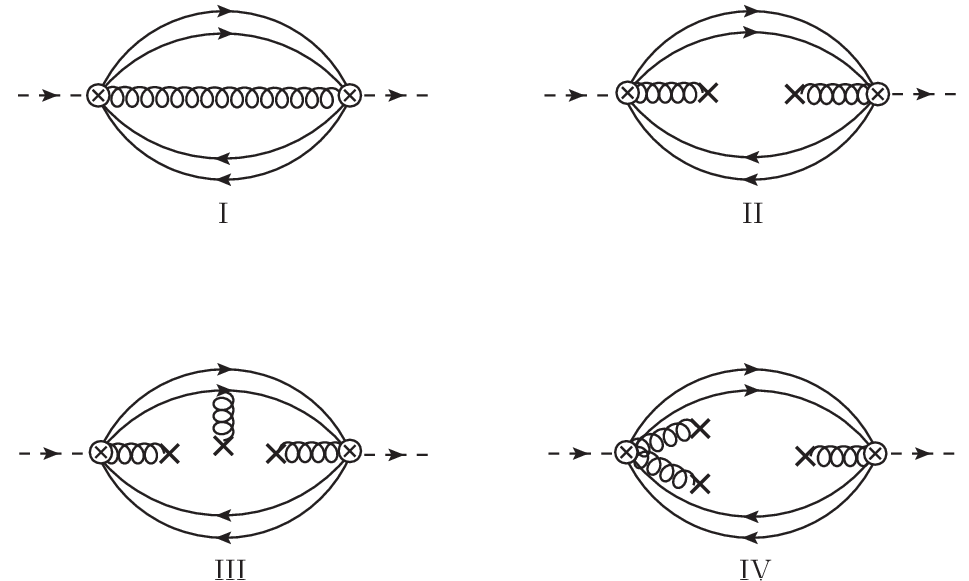}
  \caption{The leading-order Feynman diagrams for the tetracharm hybrid states are shown, with their corresponding contributions to the spectral density given in Eq.~\eqref{Pi-MB}. Only representative diagrams for each type are displayed; topologically equivalent ones are omitted for clarity. Diagrams I and II represent the perturbative contribution and the effects of the two-gluon condensate, respectively. Diagrams III and IV illustrate contributions from the three-gluon condensate.}\label{Feyn-Diag}
\end{figure}

In the mass computation of tetraquark hybrid states, the Feynman diagram expansion corresponding to Eq.~\eqref{Pi-MB}, as illustrated in Fig.~\ref{Feyn-Diag}, encompasses distinct physical processes: Diagram I embodies perturbative QCD effects, Diagram II manifests the two-gluon condensate contributions, while Diagrams III and IV correspond to the three-gluon condensate effects.

To maintain conciseness in the main text, we have selected a representative current operator for each quantum number to present. The full expressions of the spectral densities $\rho^{\text{OPE}}(s)$ corresponding to these appropriately chosen currents are provided in full in the Appendix.

From a phenomenological perspective, by isolating the ground-state contribution of tetraquark hybrid states, we obtain the hadronic representation
\begin{eqnarray}\label{Pi-hadron}
  \Pi(q^2) = \frac{(\lambda_X)^2}{(M_X)^2 - q^2} + \int_{s_0}^\infty ds \frac{\rho^h(s)}{s - q^2},
\end{eqnarray}
where $M_{X}$ marks the tetraquark hybrid state mass and $\rho^{h}(s)$ captures the spectral density, including contributions from highly excited states and continuum contribution. This phenomenological approach yields the coupling constant $\lambda_{X}$ defined as
\begin{eqnarray}
\langle 0|j^{0^{--}}(0)|X\rangle &=& \lambda_X.
\end{eqnarray}

Applying the Borel transformation to the phenomenological representation (as articulated in Eq.~\eqref{Pi-hadron}) and aligning it with Eq.~\eqref{Pi-MB}, we derive the master equation
\begin{eqnarray}\label{main-equation}
(\lambda_{X})^{2}\exp\left(-\frac{M_{X}^{2}}{M_{B}^{2}}\right)=\int_{(4 m_Q)^2}^{s_{0}} ds \rho^{\text{OPE}}(s) e^{-s/M_B^2}.
\end{eqnarray}

Ultimately, we differentiate Eq.~\eqref{main-equation} with respect to $\frac{1}{M_{B}^{2}}$ to extract the mass of the tetraquark hybrid state
\begin{eqnarray}\label{main-function}
  M_X(s_0, M_B^2) &=& \sqrt{-\frac{L_1(s_0, M_B^2)}{L_0(s_0, M_B^2)}},
\end{eqnarray}
wherein the moments $L_{0}$ and $L_{1}$ are respectively defined as
\begin{eqnarray}\label{OPE-function}
  L_0(s_0, M_B^2) &=& \int_{(4 m_Q)^2}^{s_{0}} ds \rho^{\text{OPE}}(s) e^{-s/M_B^2}, \label{L0} \\
  L_1(s_0, M_B^2) &=& \frac{\partial}{\partial (M_B^2)^{-1}} L_0(s_0, M_B^2).
\end{eqnarray}
\section{Numerical analysis}
For the numerical analysis in this section, we adopt the following parameter values for quark masses and gluon condensates~\cite{Shifman:1978bx, Reinders:1984sr, Narison:1989aq, Colangelo:2000dp, Shifman:1978by}
\begin{eqnarray}
  m_c (m_c) = \overline{m}_c&=& (1.27 \pm 0.03) \; \text{GeV},\nonumber\\
  m_b (m_b) = \overline{m}_b&=& (4.18 \pm 0.03) \; \text{GeV},\nonumber\\
  \langle g_{s}^{2}G^{2}\rangle&=&0.88 \; \text{GeV}^{4},\nonumber\\
  \langle g_{s}^{3}G^{3}\rangle&=&0.045 \; \text{GeV}^{6},
\end{eqnarray}
where $m_{c}$ and $m_{b}$ represent the running masses of charm and bottom quarks respectively in the $\overline{MS}$ renormalization scheme.

Within the framework of QCD sum rules, the mass determination of tetraquark hybrid states hinges critically on the judicious selection of both the continuum threshold parameter $s_{0}$ and the Borel parameter $M_{B}^{2}$. The establishment of viable sum rules mandates that these parameters satisfy dual fundamental criteria~\cite{Shifman:1978bx, Shifman:1978by, Reinders:1984sr, Colangelo:2000dp}: Primarily, to ensure a physically meaningful description of the ground state hadron based on truncated operator product expansion (OPE), the convergence of the OPE series must be rigorously guaranteed, thereby precluding substantial errors induced by higher dimensional term truncations. In practical implementation, we delineate the reliable working window for $M_{B}^{2}$ by systematically comparing the relative contributions of successive higher-dimensional condensates to the total OPE contribution. This convergence criterion admits quantitative formulation as follows
\begin{eqnarray}
  R_{i}^{\text{cond}}(s_0, M_B^2) = \frac{L_0^{\text{dim}}(s_0, M_B^2)}{L_0(s_0, M_B^2)}\, ,
\end{eqnarray}
here, the subscript $i$ denotes distinct currents, while the superscript ``dim'' explicitly specifies the dimensional attribute of condensate terms in the OPE. Building upon this foundation, for a given continuum threshold $s_0$, we systematically determine the lower bound $(M_{B}^{2})_{\text{min}}$ of the Borel parameter $M_B^2$ through rigorous analysis. In practical implementation, to strictly ensure OPE series convergence, the primary validation criterion requires that the contribution from the three-gluon condensate $\langle g_s^3G^3\rangle$ does not exceed 10\% of the total OPE contribution. This convergence criterion is established in accordance with standard QCD sum rule practices. Furthermore, to guarantee that the mass equation is dominated by ground-state signals, the pole contribution (PC) must satisfy a significance threshold. Specifically, following the criteria outlined in references~\cite{Colangelo:2000dp,Matheus:2006xi}, the PC must exceed 40\%, mathematically expressed as
\begin{eqnarray}
  R_{i}^{\text{PC}}(s_0, M_B^2) = \frac{L_0(s_0, M_B^2)}{L_0(\infty, M_B^2)} \; , \label{RatioPC}
\end{eqnarray}
under this constraint, contributions from higher excited states and continuum are substantially suppressed. This requirement precipitates the determination of a critical $M_{B}^{2}$ value serving as the upper threshold $(M_{B}^{2})_{\text{max}}$.

Within the rigorous theoretical framework of QCD sum rules, the determination of the continuum threshold parameter $s_0$ must satisfy exacting physical constraints, particularly requiring the tetracharm hybrid state mass $M_X$ to demonstrate negligible parametric dependence on $s_0$. Building upon the established methodology in references~\cite{Finazzo:2011he, Qiao:2013raa, Qiao:2013dda}, our analysis systematically optimizes $s_0$ through an exhaustive procedure that begins with a comprehensive scan across the $\sqrt{s_0}$ parameter space to identify regions satisfying fundamental Borel platform criterion, subsequently pinpointing the specific $\sqrt{s_0}$ value that produces maximally stable Borel mass platform, and while maintaining the essential physical scaling relation $\sqrt{s_0} \approx (M_X + \delta)$ where the empirical offset $\delta$ resides within 0.40-0.80 GeV. The ultimate selection of the central $\sqrt{s_0}$ value, derived from this optimized stability platform, incorporates systematic uncertainties through symmetric variations of $\Delta\sqrt{s_0} = \pm 0.2$ GeV as prescribed in the foundational literature.

\begin{figure}[htb]
\begin{center}
\includegraphics[width=6.8cm]{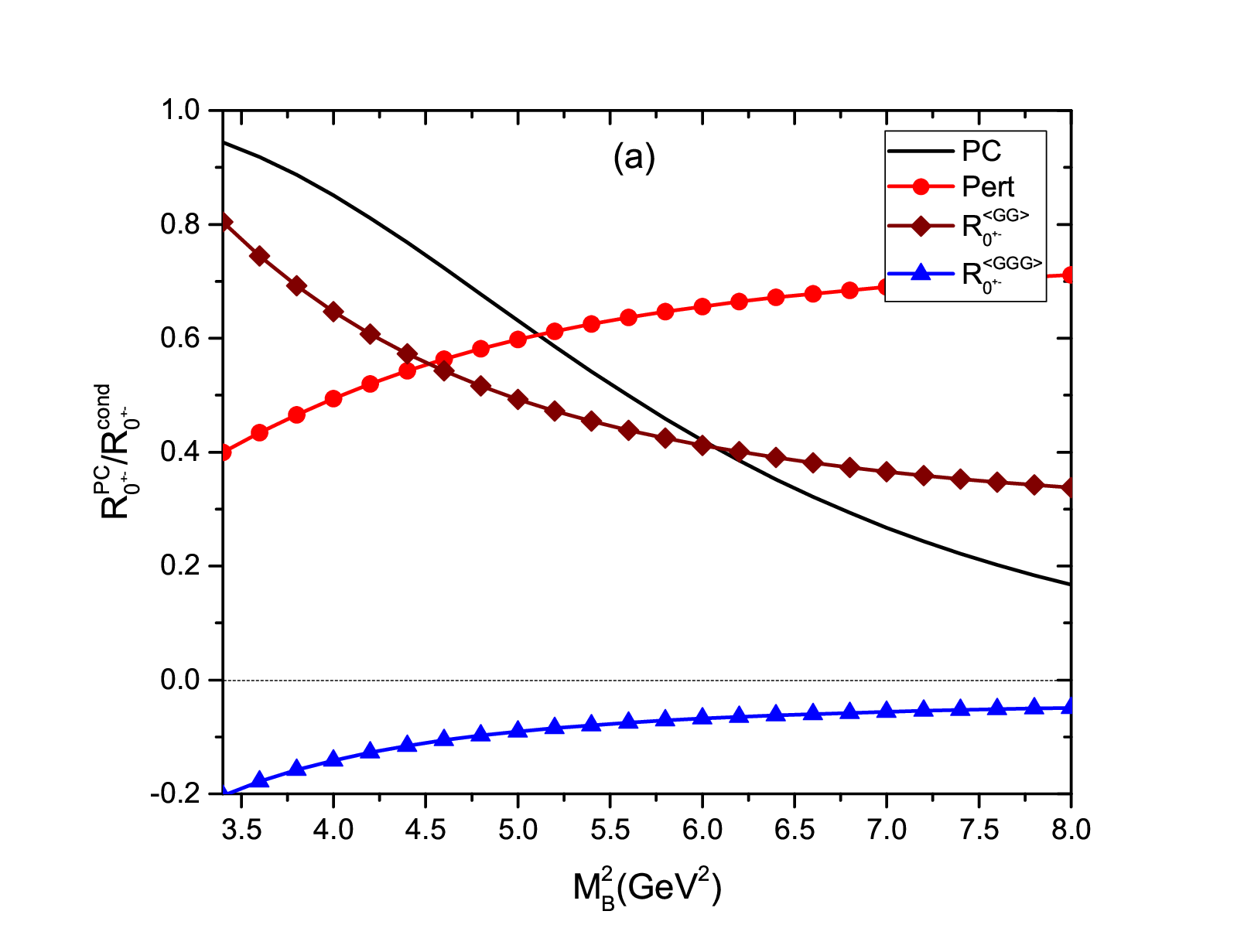}
\includegraphics[width=6.8cm]{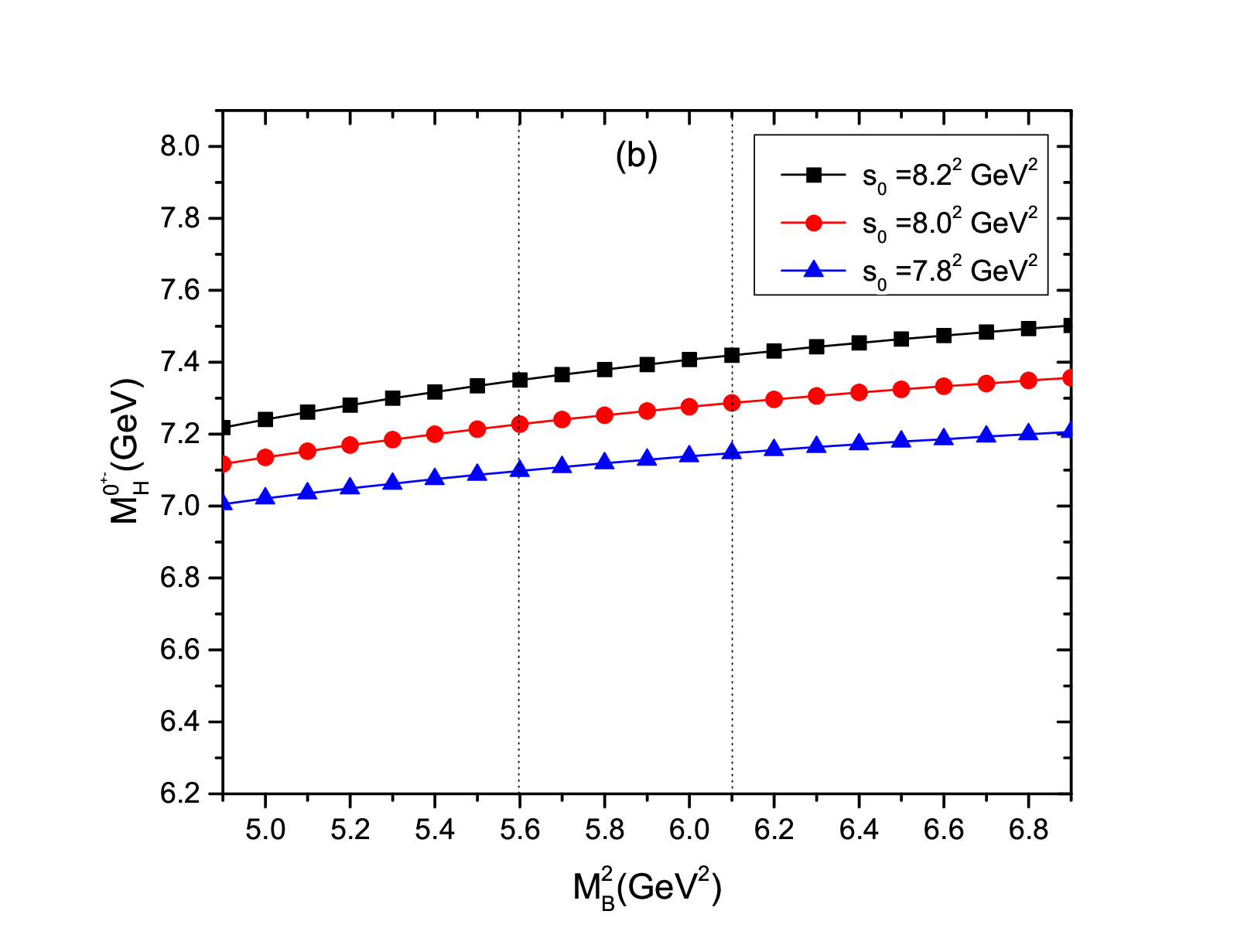}
\caption{(color).The numerical analysis results for current operator $j^{0^{+-}}$ are presented graphically: Figure (a) displays the variation of pole contribution ratio $R^{\text{PC}}_{0^{+-}}$ and OPE convergence ratio $R^{\text{cond}}_{0^{+-}}$ with respect to Borel parameter $M_B^2$ at the central value of continuum threshold $s_0$, while Figure (b) illustrates the evolution of state mass $M_{\text{H}}^{0^{+-}}$ versus $M_B^2$, where three distinct curves correspond to parameter choices of $s_0=7.8^{2}\, \text{GeV}^{2}$ (lower bound), $8.0^{2}\, \text{GeV}^{2}$ (central value), and $8.2^{2}\,  \text{GeV}^{2}$ (upper bound). The valid Borel working window, demarcated by vertical gray dashed lines in the figures, is established relative to the central $s_0$ value.}
\label{fig2}
\includegraphics[width=6.8cm]{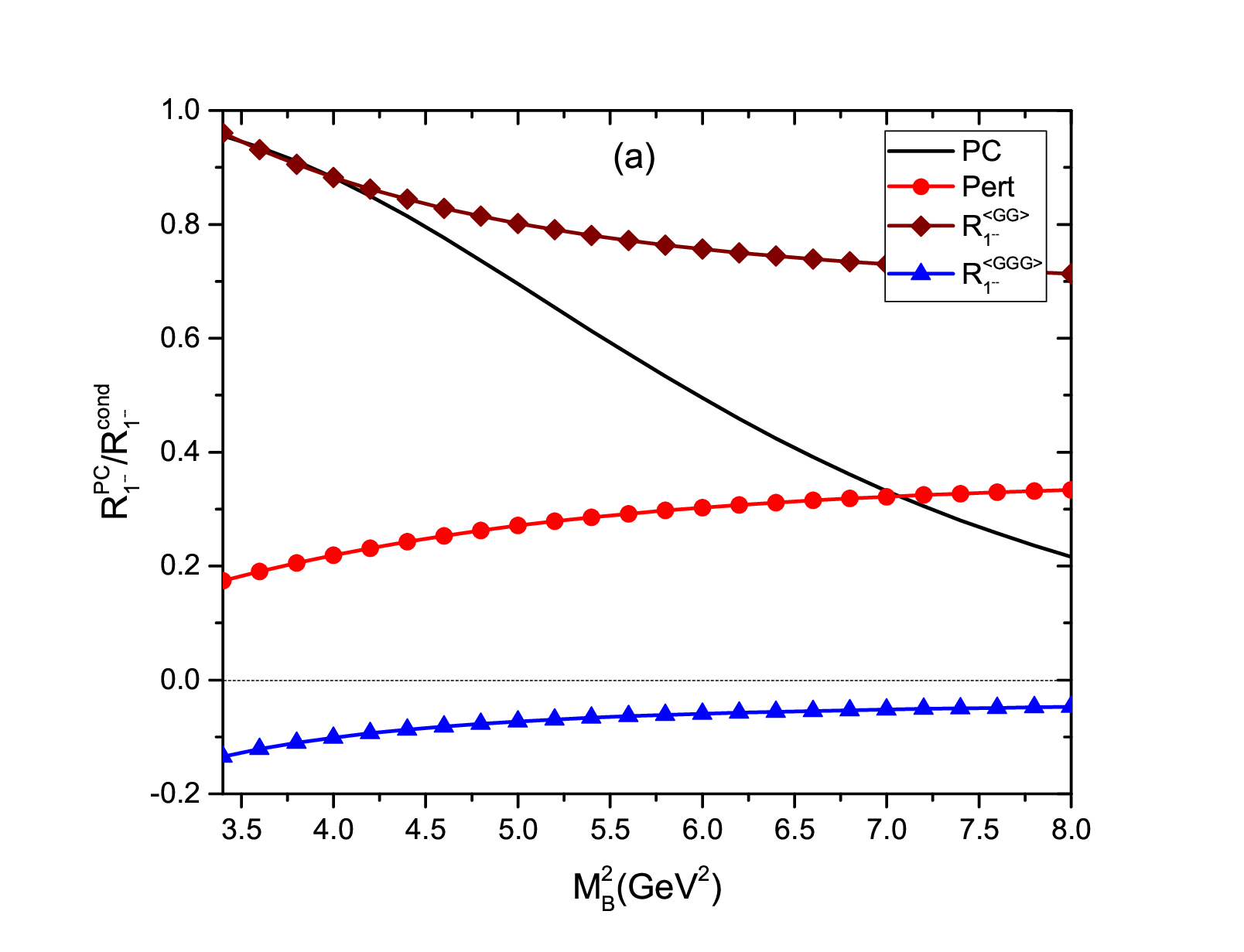}
\includegraphics[width=6.8cm]{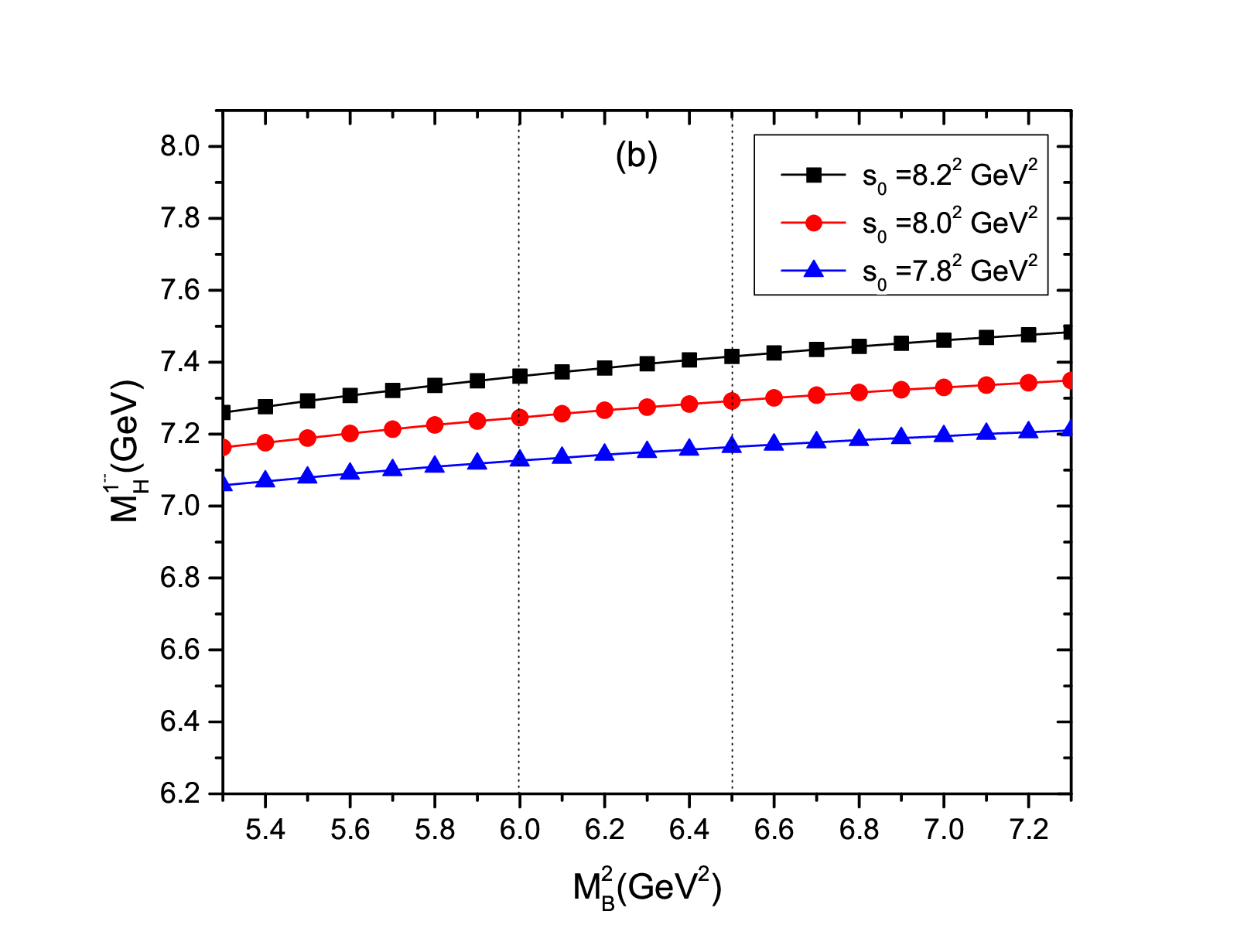}
\caption{(color). Following the same caption format as Fig.~\ref{fig2} but now addressing the current $j_{A}^{1^{--}}$.}
\label{fig3}
\end{center}
\end{figure}

\begin{figure}
\begin{center}
\includegraphics[width=6.8cm]{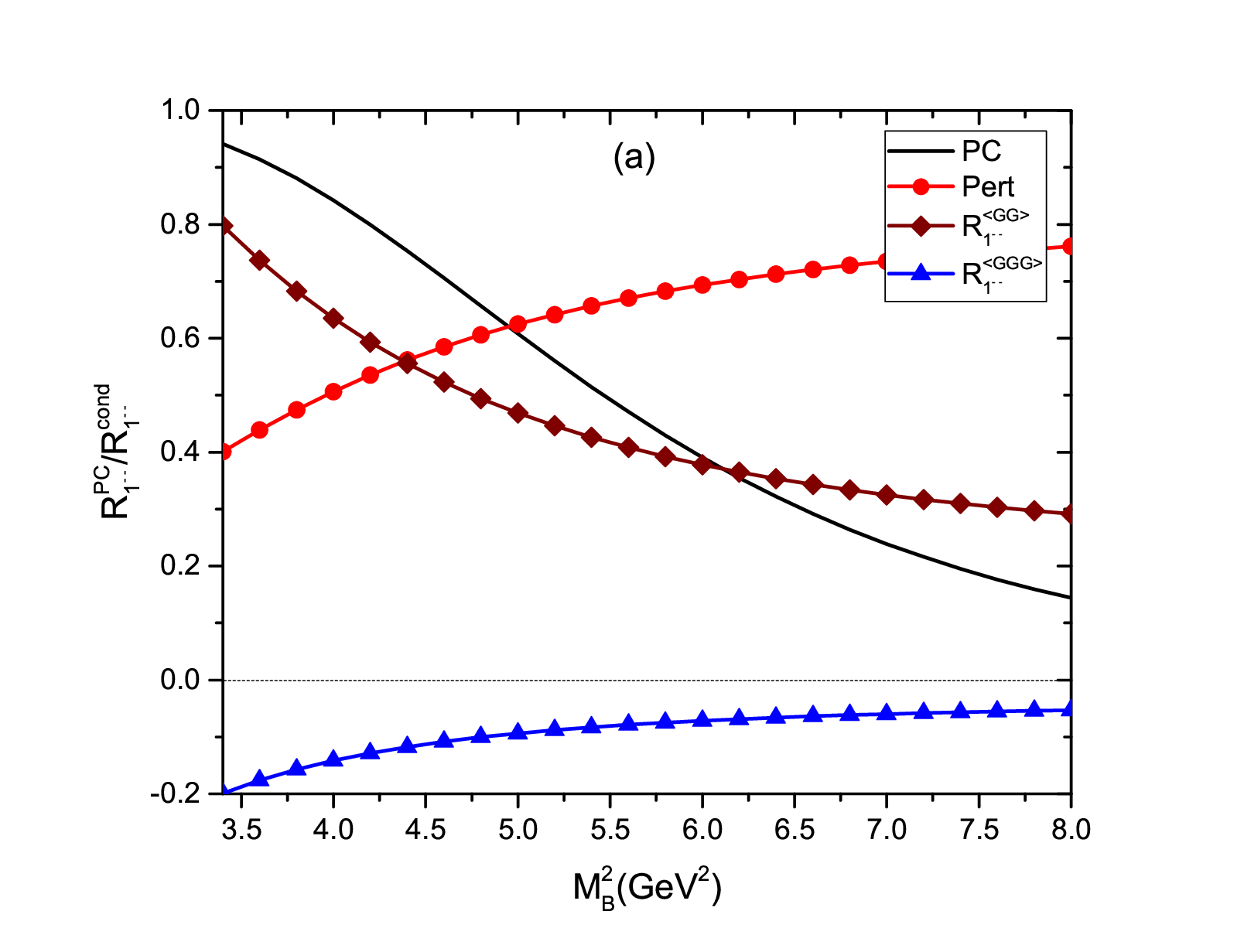}
\includegraphics[width=6.8cm]{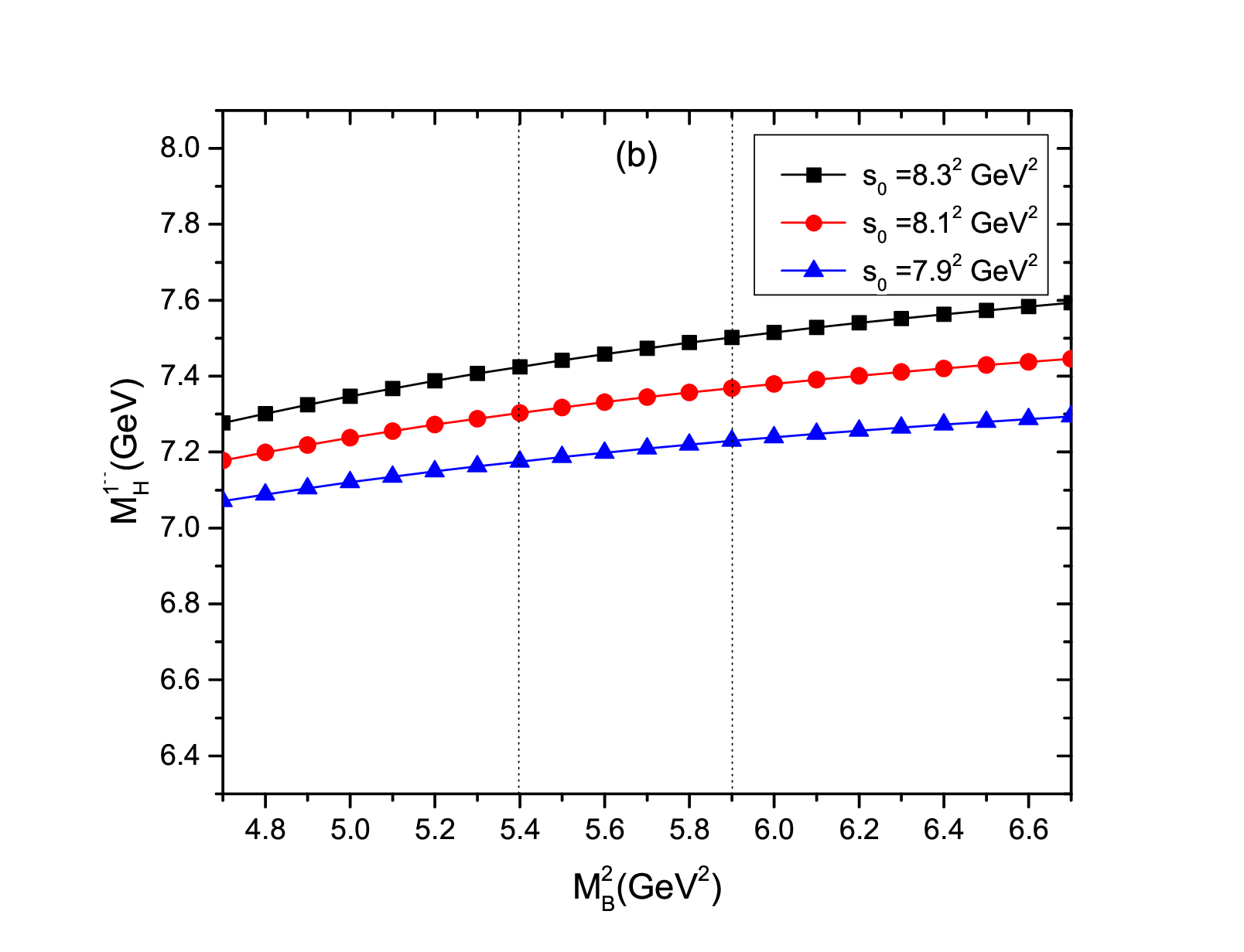}
\caption{(color). Following the same caption convention as Fig.~\ref{fig2} but now pertaining to the current $j_{C}^{1^{--}}$.}
\label{fig4}
\includegraphics[width=6.8cm]{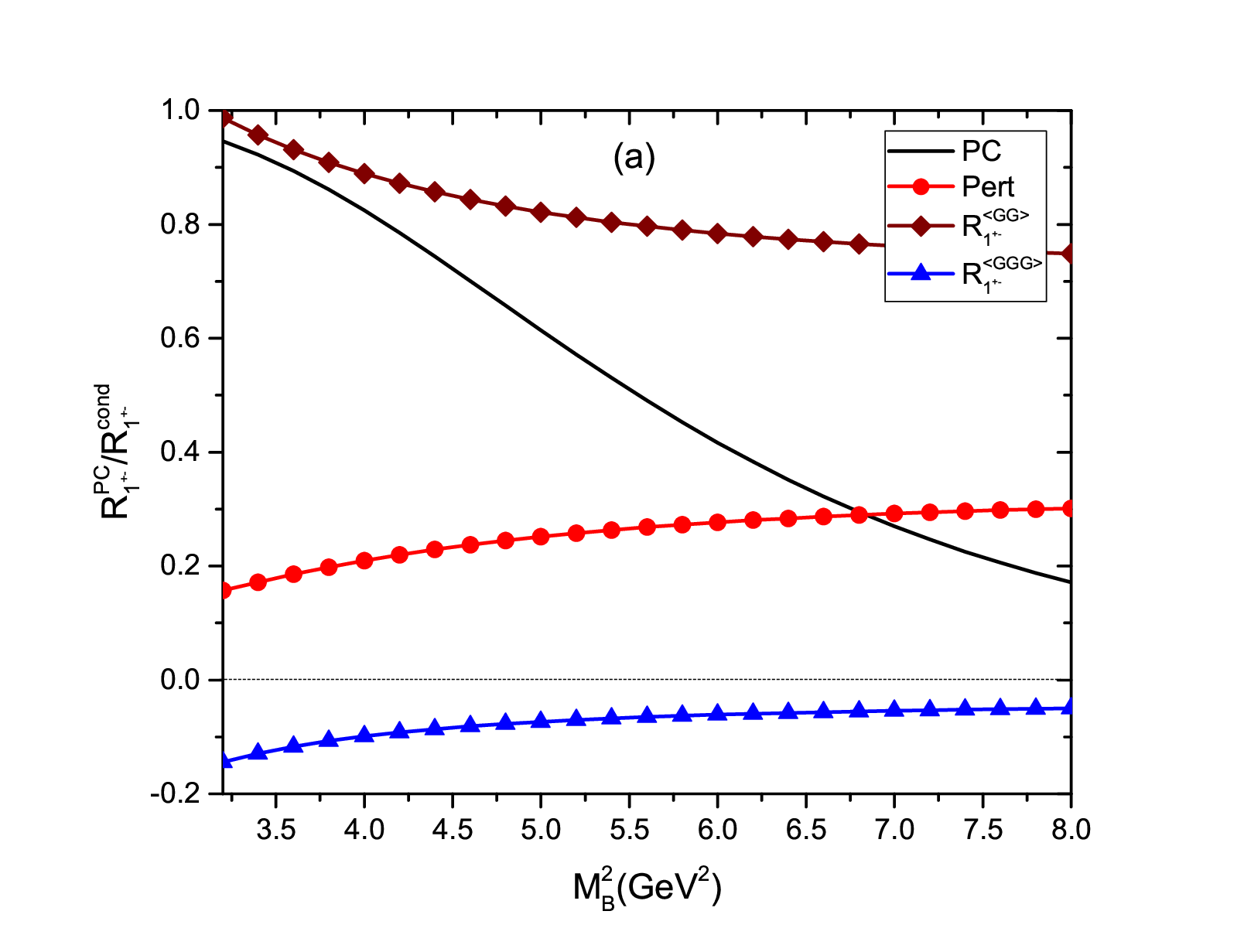}
\includegraphics[width=6.8cm]{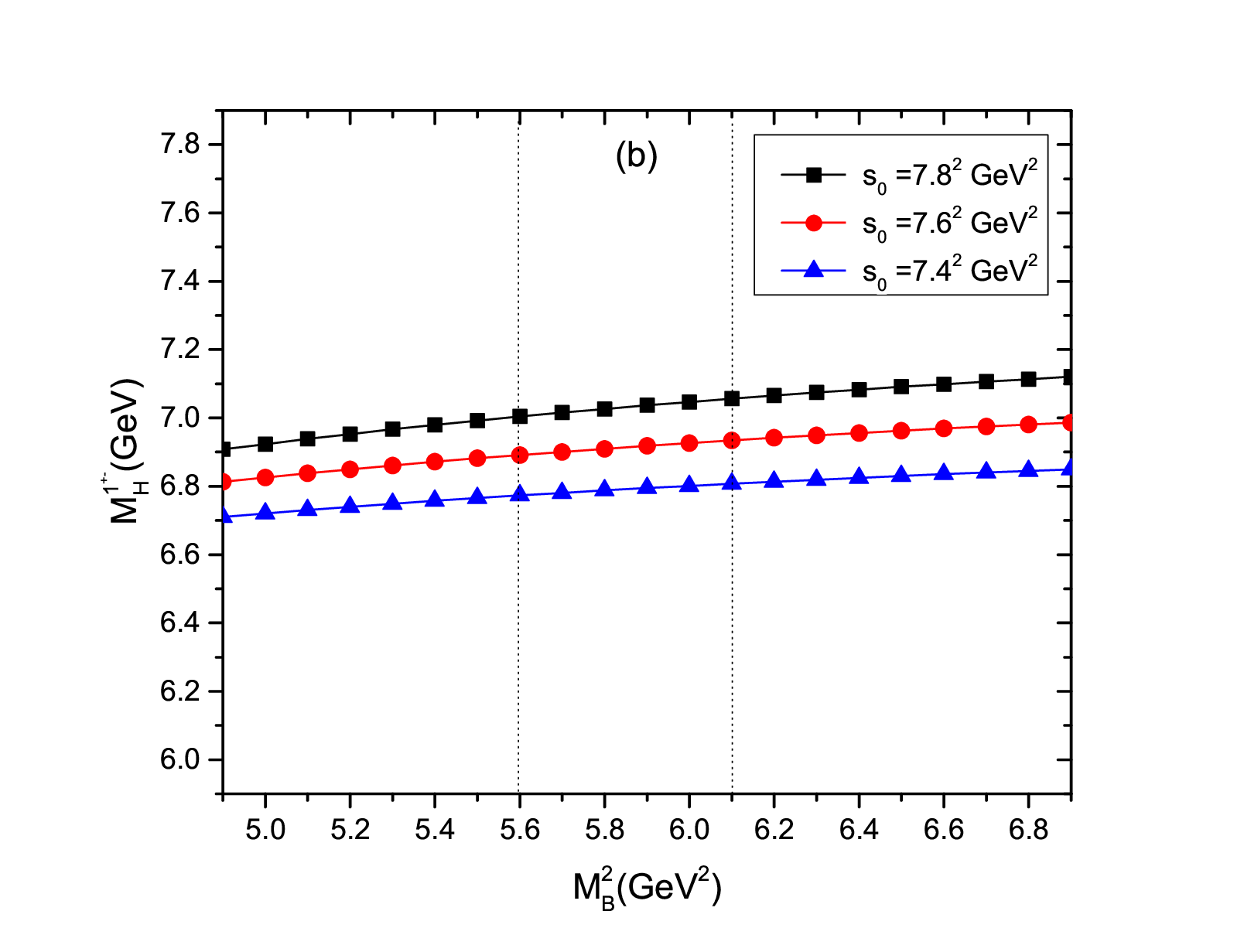}
\caption{(color). Following the same caption convention as Fig.~\ref{fig2} but now pertaining to the current $j_{B}^{1^{+-}}$.}
\label{fig5}
\includegraphics[width=6.8cm]{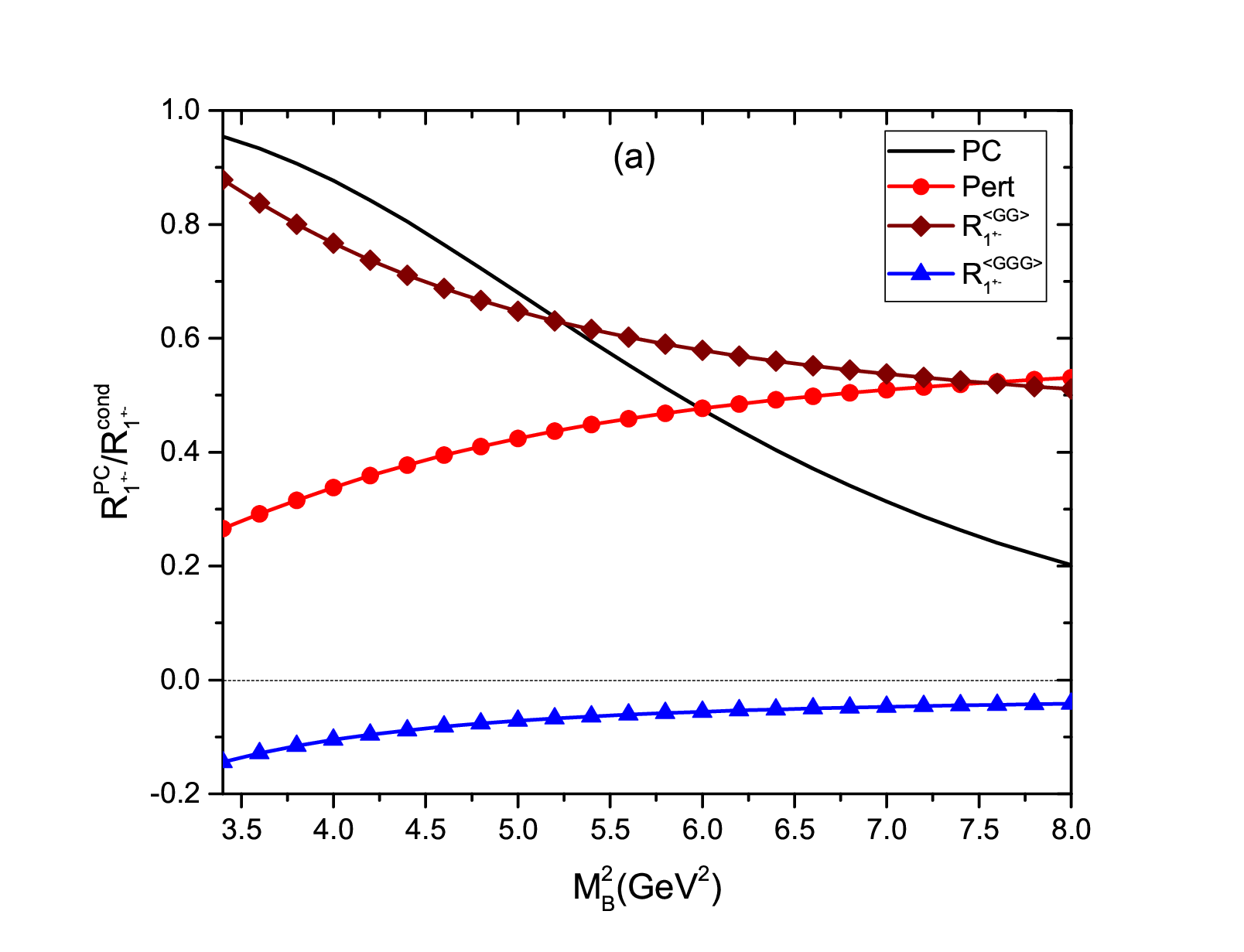}
\includegraphics[width=6.8cm]{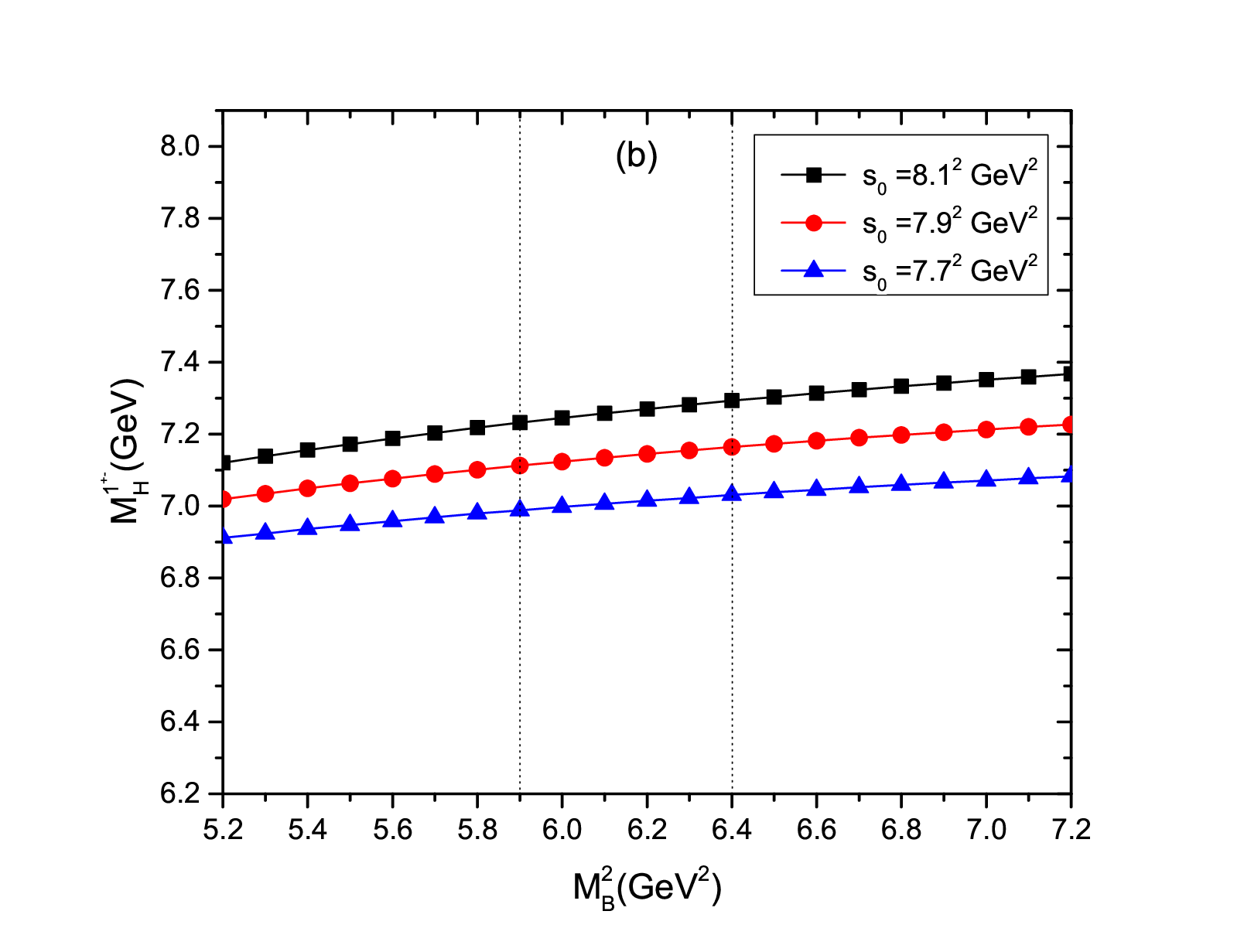}
\caption{(color). Following the same caption convention as Fig.~\ref{fig2} but now pertaining to the current $j_{D}^{1^{+-}}$.}
\label{fig6}
\end{center}
\end{figure}

\begin{figure}
\begin{center}
\includegraphics[width=6.8cm]{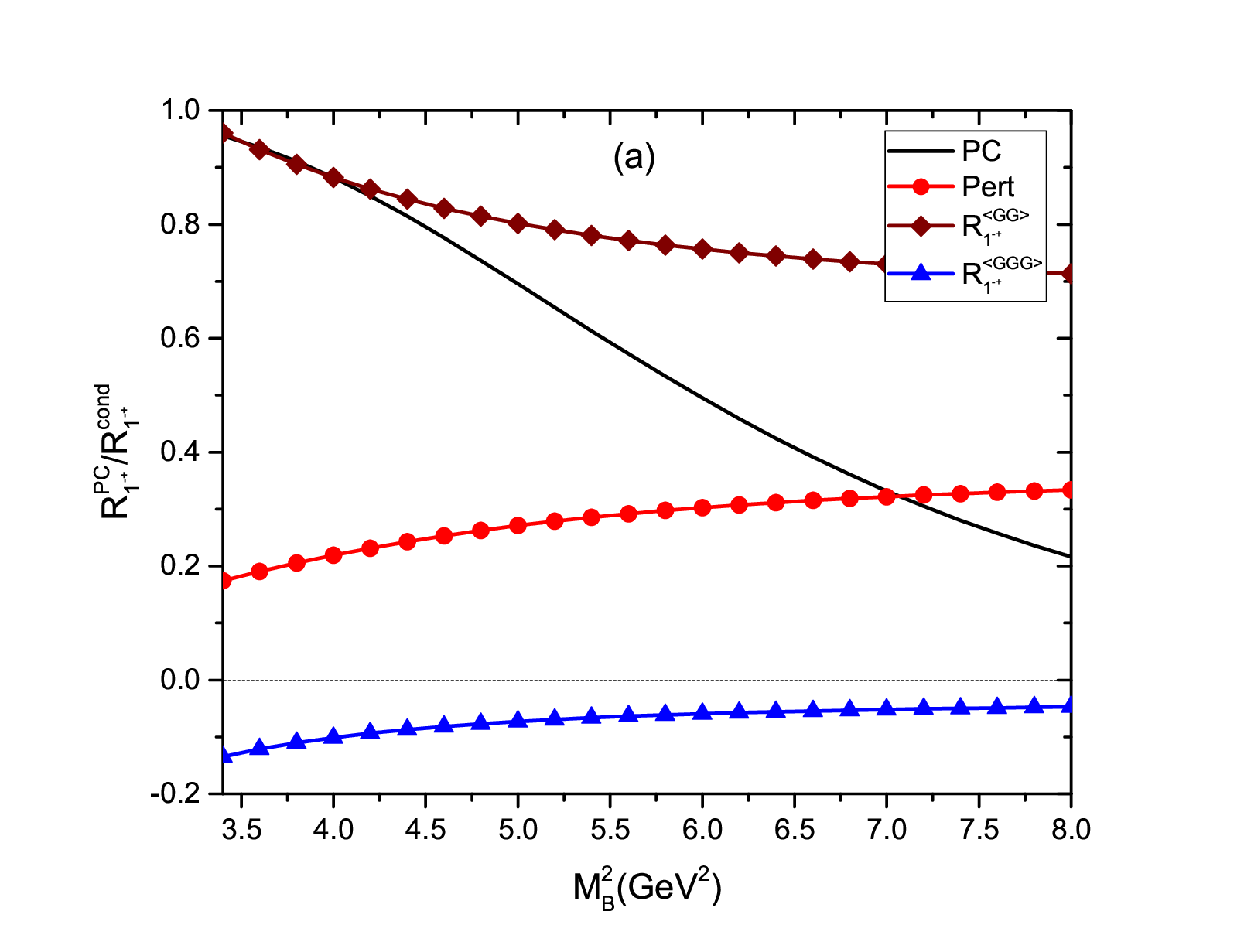}
\includegraphics[width=6.8cm]{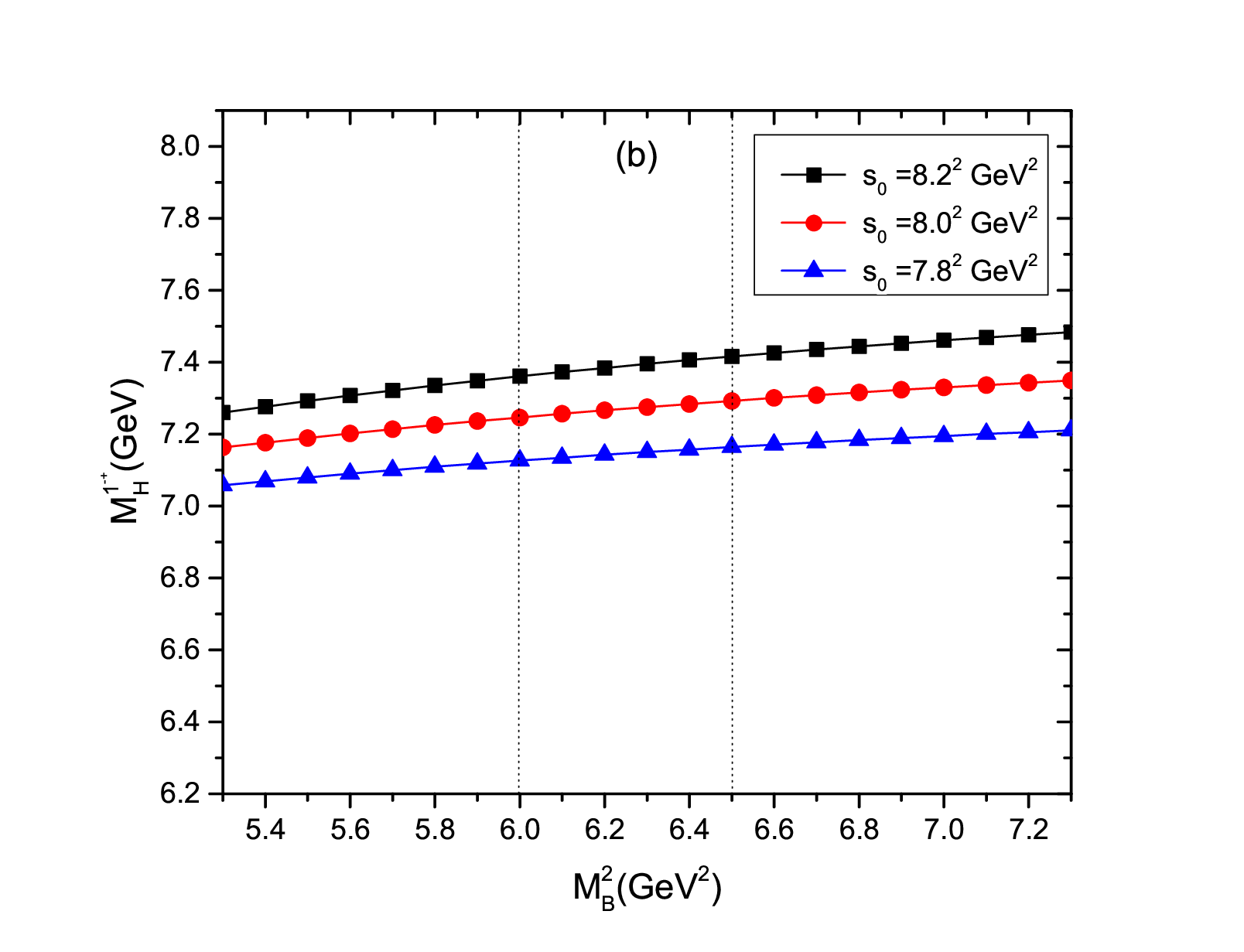}
\caption{(color). Following the same caption convention as Fig.~\ref{fig2} but now pertaining to the current $j_{A}^{1^{-+}}$.}.
\label{fig7}
\includegraphics[width=6.8cm]{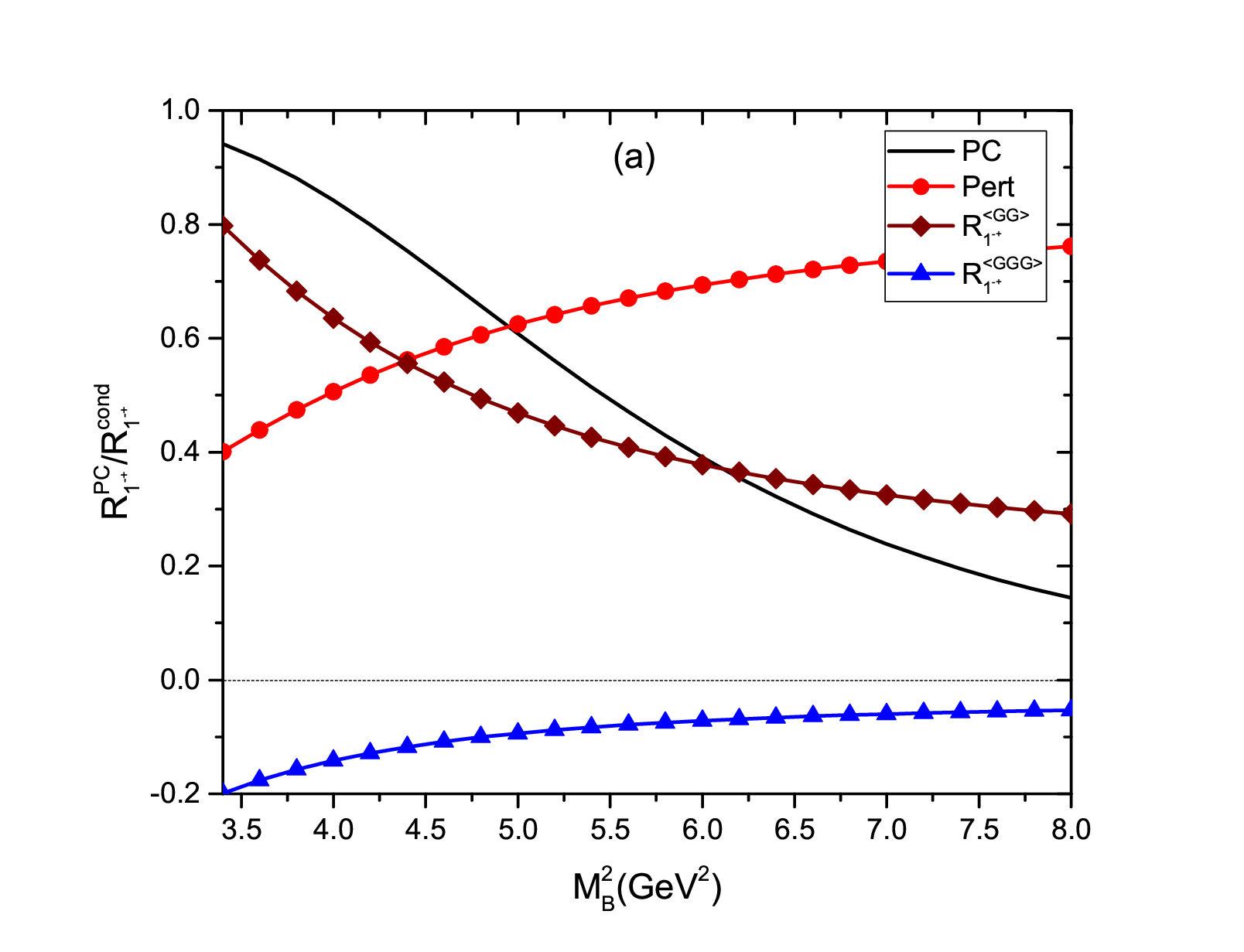}
\includegraphics[width=6.8cm]{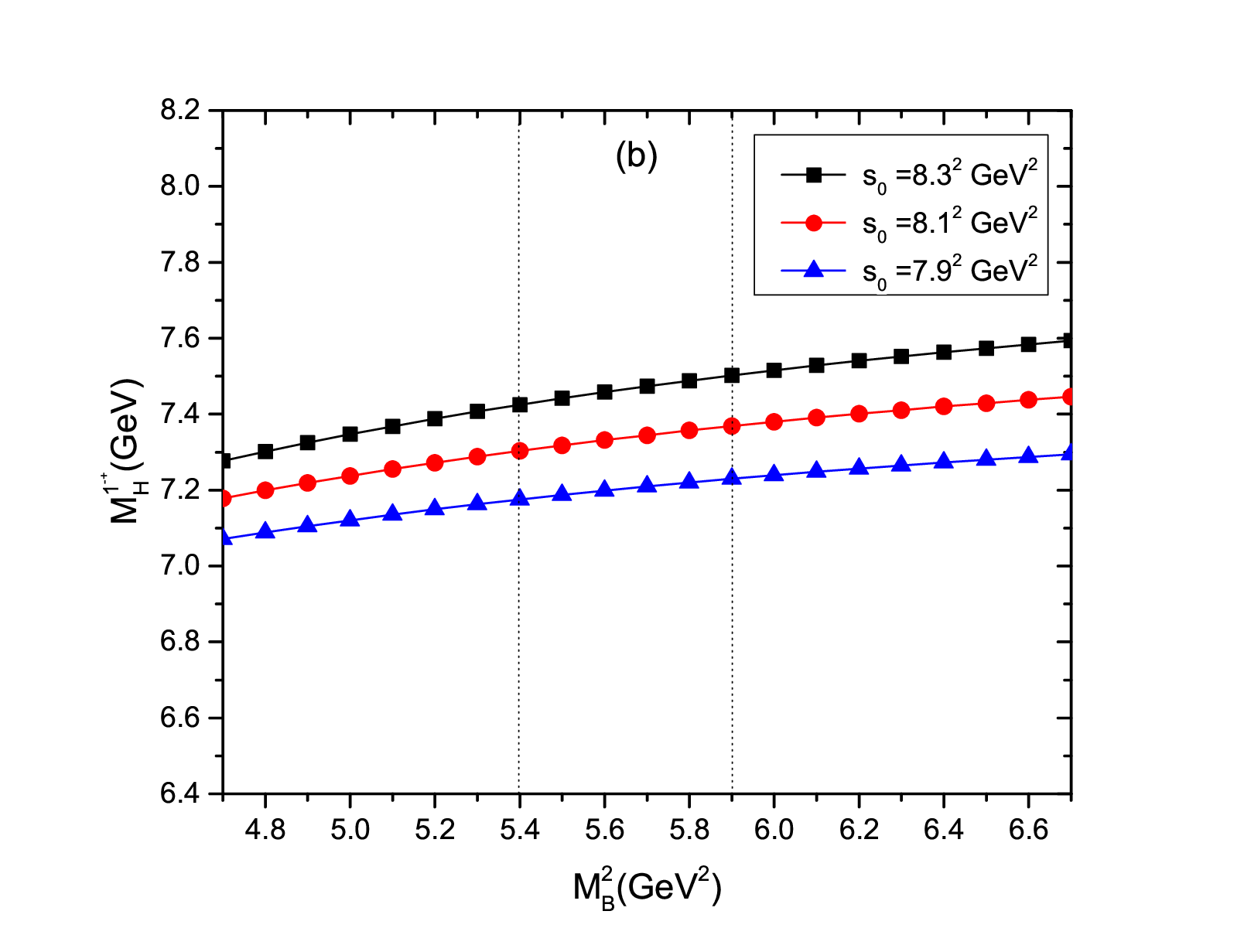}
\caption{(color). Following the same caption convention as Fig.~\ref{fig2} but now pertaining to the current $j_{C}^{1^{-+}}$.}
\label{fig8}
\includegraphics[width=6.8cm]{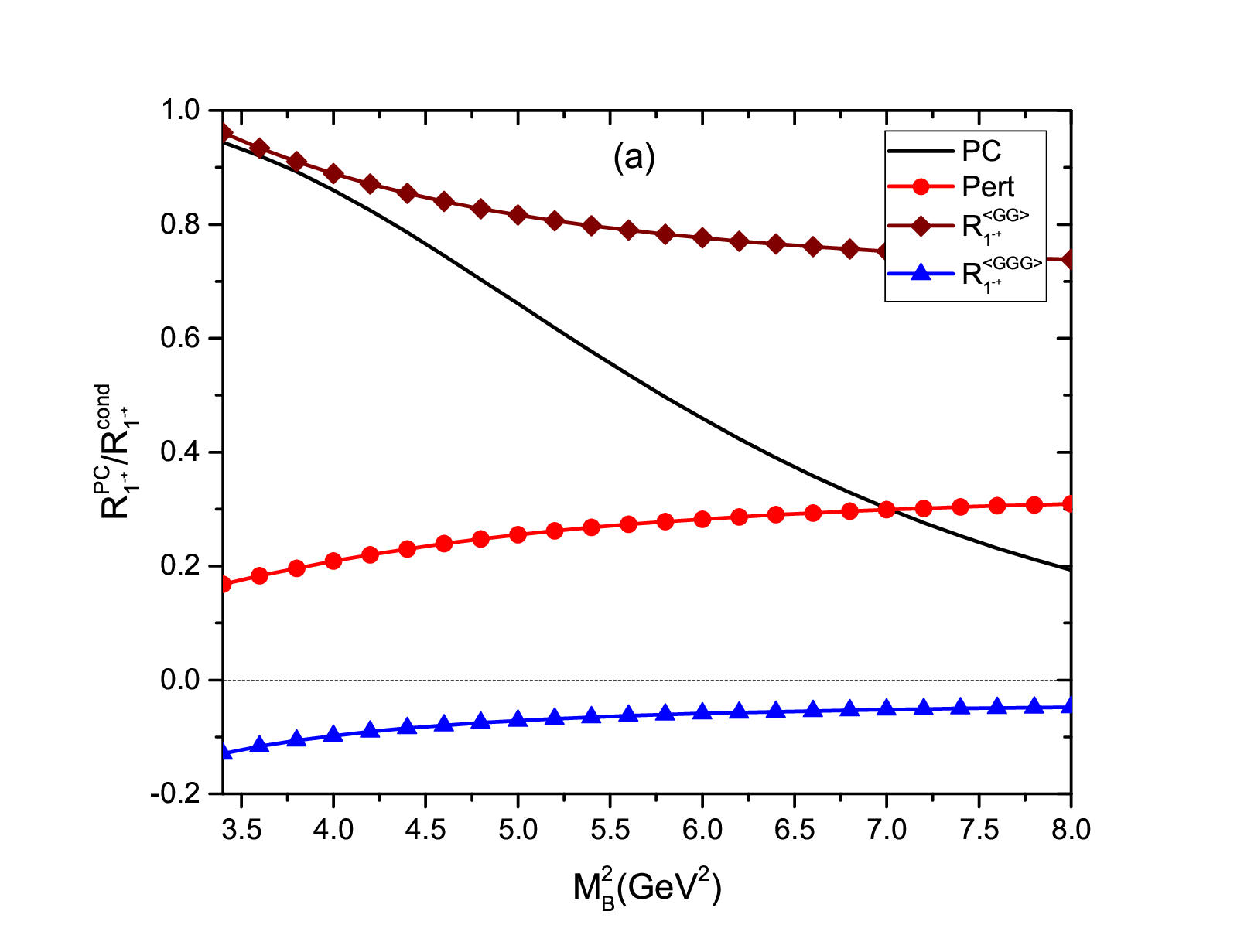}
\includegraphics[width=6.8cm]{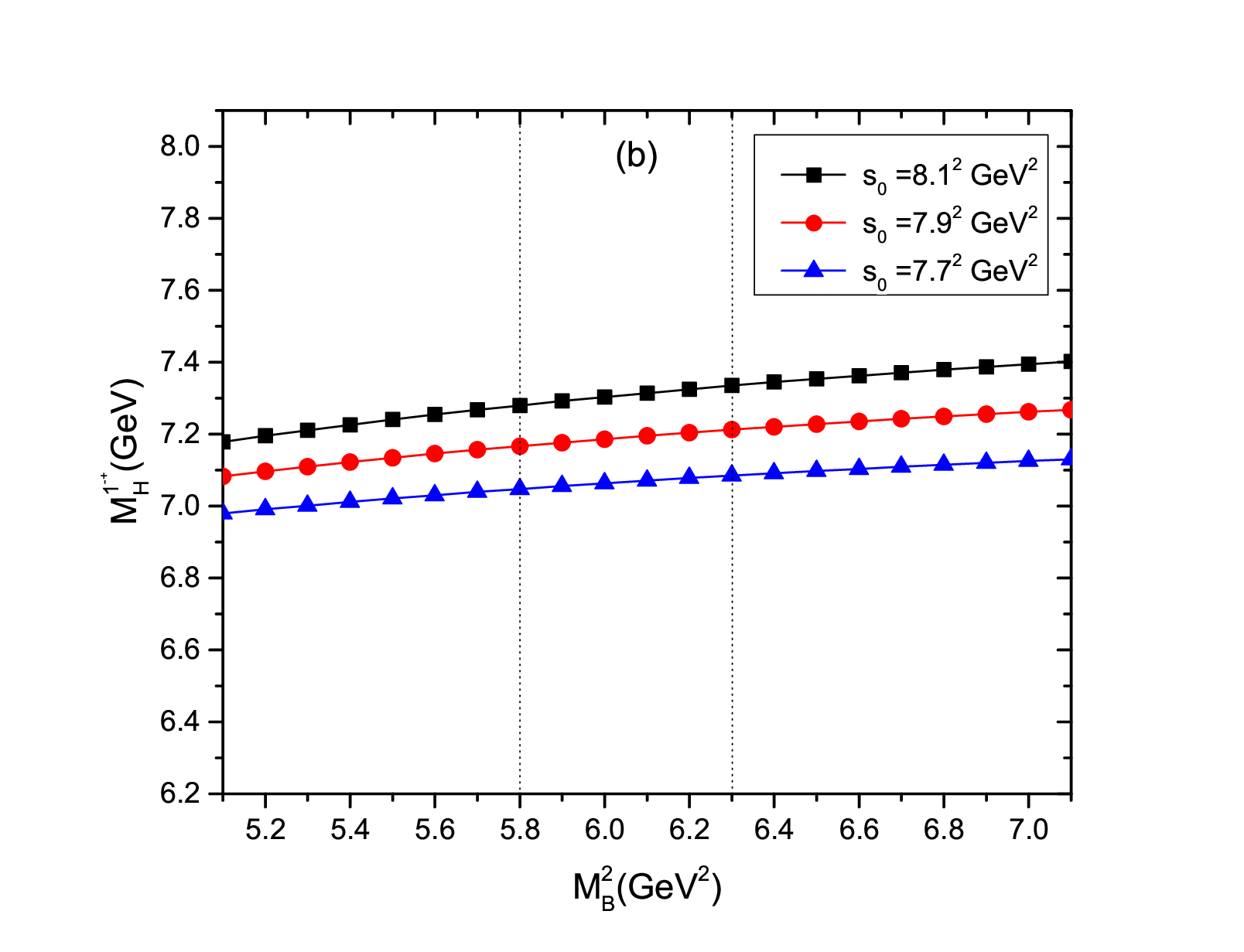}
\caption{(color). Following the same caption convention as Fig.~\ref{fig2} but now pertaining to the current $j_{D}^{1^{-+}}$.}
\label{fig9}
\end{center}
\end{figure}

\begin{figure}[htb]
\begin{center}
\includegraphics[width=6.8cm]{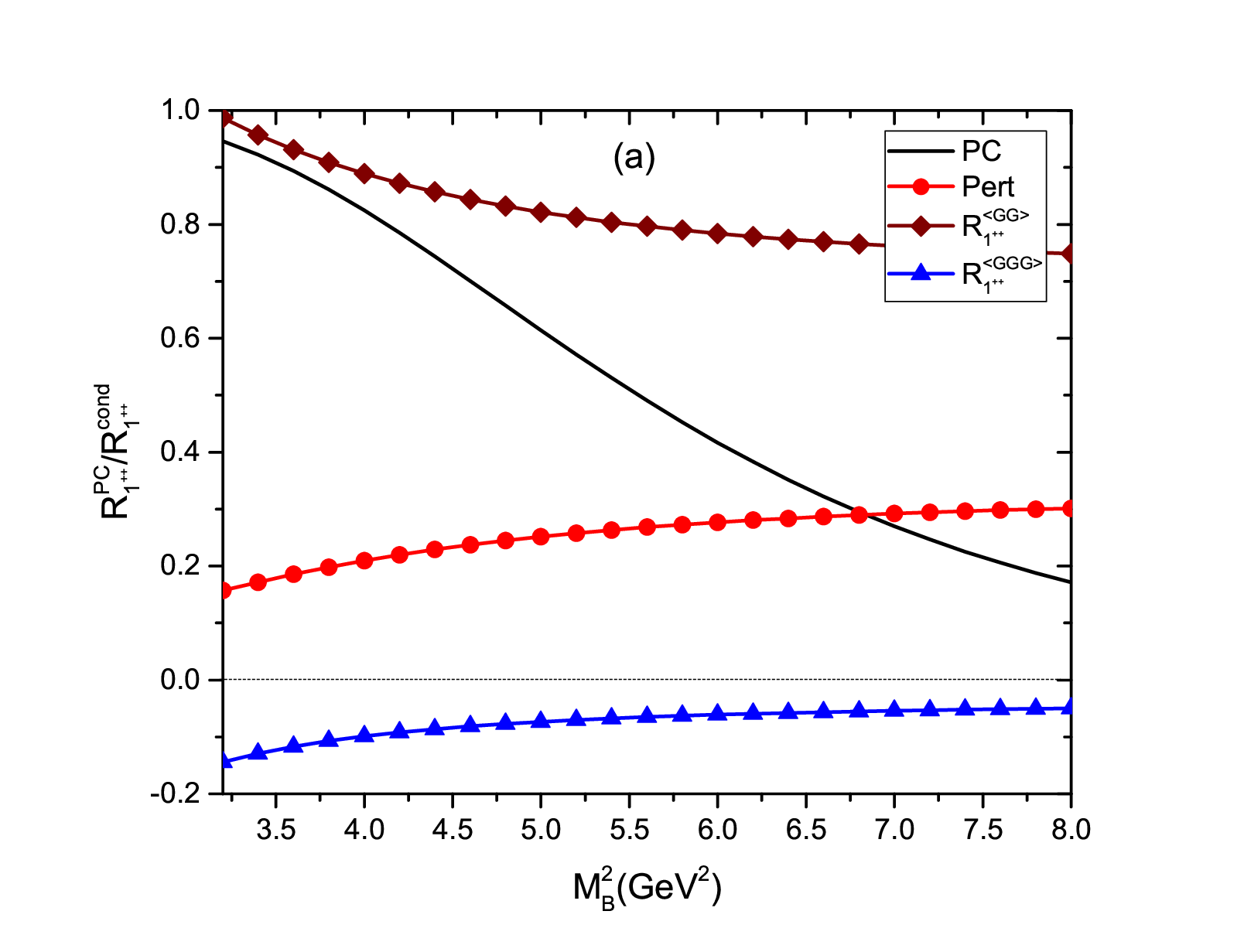}
\includegraphics[width=6.8cm]{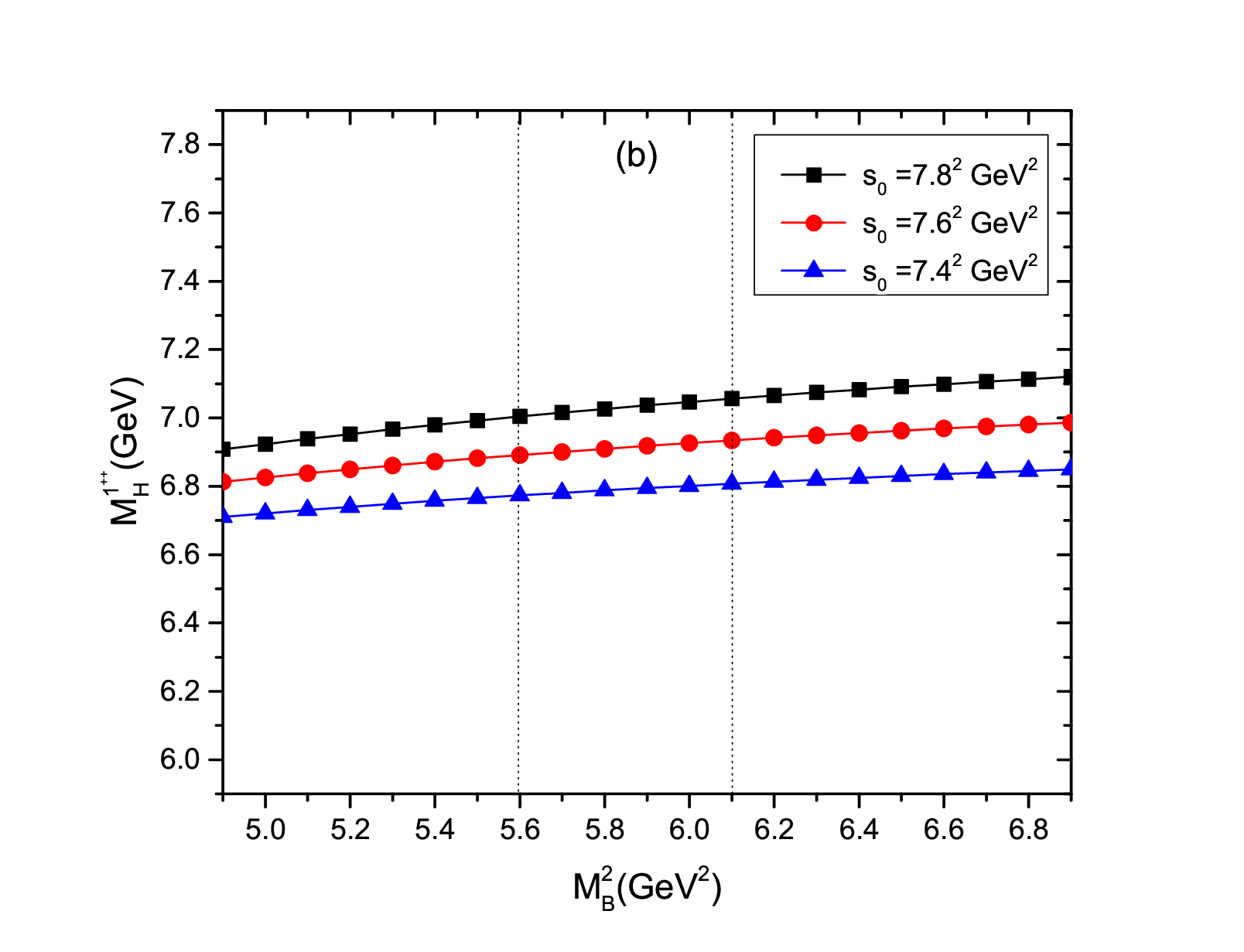}
\caption{(color). Following the same caption convention as Fig.~\ref{fig2} but now pertaining to the current $j_{B}^{1^{++}}$.}
\label{fig10}
\end{center}
\end{figure}

Based on the established theoretical framework and optimized parameter selection, we systematically investigated the mass spectrum characteristics of tetraquark hybrid states. For each quantum number configuration, we selected the most representative current operator and presented in detail: the mass variation curves as functions of the Borel parameter $M_B^2$, quantitative analysis results of pole contribution (PC), and diagrams demonstrating the convergence of the OPE series. Taking current $j^{0^{+-}}(x)$ in Fig.~\ref{fig2} as exemplar, Fig.~\ref{fig2}(a) displays the evolution of ratios $R^{\text{PC}}_{0^{+-}}$ and $R^{\text{cond}}_{0^{+-}}$ versus $M_{B}^{2}$ at fixed $s_{0} = 8.0^{2} \, \text{GeV}^{2}$, while Fig.~\ref{fig2}(b) charts the mass dependence on $M_{B}^{2}$ across different $s_{0}$ values. The vertical lines in Fig.~\ref{fig2}(b) demarcate the upper and lower bounds of the valid Borel window at central $s_{0}$. Parallel analyses apply to Figs.~\eqref{fig3}-\eqref{fig10}.

Drawing upon the three fundamental criteria of QCD sum rules, Table \ref{tab2} systematically compiles the predicted mass values of tetracharm hybrid states alongside critical parameters including the corresponding Borel parameters, continuum thresholds, pole contributions, and pole residues. The analysis reveals that within the optimized Borel window, the pole dominance condition is rigorously satisfied, with the mass curves exhibiting excellent platform characteristics. The central values correspond to computational results in the most stable region of $M_{B}^{2}$, while the error margins comprehensively incorporate variations from condensate parameters, quark masses, and the $s_{0}$ and $M_{B}^{2}$ parameter ranges. Notably, apart from the current operators enumerated in Table \ref{tab2}, alternative current configurations were excluded due to their failure to yield positive-definite spectral densities---a definitive indication that these current structures cannot support stable tetracharm hybrid state formation.
\begin{table}[htb]
\begin{center}
\begin{tabular}{|c|c|c|c|c|c|c|c|c}\hline\hline
 $J^{PC}$& $M_B^2\,(\rm{GeV}^2)$  & $\sqrt{s_{0}}\,(\rm{GeV})$ & PC & $M_{X}$ $(\rm{GeV})$ & $\lambda_{X}$ $(\rm{10^{-1}GeV^{5}} )$\\ \hline
 $j^{0^{+-}}$ & $5.60\!-\!6.10$ & $8.00\pm 0.20$  & $(50\!-\!40)\%$ & $7.26_{-0.16}^{+0.16}$  & $0.75_{-0.07}^{+0.06}$  \\ \hline
 $j_{A}^{1^{--}}$ & $6.00\!-\!6.50$ & $8.00\pm 0.20$  & $(50\!-\!40)\%$ & $7.27_{-0.14}^{+0.15}$  & $0.38_{-0.04}^{+0.03}$  \\ \hline
 $j_{C}^{1^{--}}$ & $5.40\!-\!5.90$ & $8.10\pm 0.20$  & $(50\!-\!40)\%$ & $7.34_{-0.17}^{+0.16}$  & $0.18_{-0.02}^{+0.02}$  \\ \hline
 $j_{B}^{1^{+-}}$ & $5.60\!-\!6.10$ & $7.60\pm 0.20$  & $(50\!-\!40)\%$ & $6.91_{-0.14}^{+0.15}$  & $0.38_{-0.04}^{+0.03}$  \\ \hline
 $j_{D}^{1^{+-}}$ & $5.90\!-\!6.40$ & $7.90\pm 0.20$  & $(50\!-\!40)\%$ & $7.14_{-0.15}^{+0.15}$  & $0.37_{-0.03}^{+0.03}$  \\ \hline
 $j_{A}^{1^{-+}}$ & $6.00\!-\!6.50$ & $8.00\pm 0.20$  & $(50\!-\!40)\%$ & $7.27_{-0.14}^{+0.15}$  & $0.28_{-0.02}^{+0.02}$  \\ \hline
 $j_{C}^{1^{-+}}$ & $5.40\!-\!5.90$ & $8.10\pm 0.20$  & $(50\!-\!40)\%$ & $7.34_{-0.16}^{+0.16}$  & $0.24_{-0.02}^{+0.02}$  \\ \hline
 $j_{D}^{1^{-+}}$ & $5.80\!-\!6.30$ & $7.90\pm 0.20$  & $(50\!-\!40)\%$ & $7.19_{-0.14}^{+0.15}$  & $0.33_{-0.03}^{+0.03}$  \\ \hline
 $j_{B}^{1^{++}}$ & $5.60\!-\!6.10$ & $7.60\pm 0.20$  & $(50\!-\!40)\%$ & $6.92_{-0.15}^{+0.14}$  & $0.29_{-0.03}^{+0.02}$  \\ \hline
 \hline
\end{tabular}
\end{center}
\caption{The ranges of Borel parameter $M_{B}^{2}$, threshold parameter $s_{0}$, pole contribution (PC), mass, and ground state pole residue for tetracharm hybrid states.}
\label{tab2}
\end{table}

Similarly, by replacing the charm quark with the bottom quark in the aforementioned theoretical calculations and numerical analyses, we can predict the masses of tetrabottom hybrid states as shown in Table \ref{tab3}.
\begin{table}[htb]
\begin{center}
\begin{tabular}{|>{\centering}p{2cm}|>{\centering}p{2cm}||>{\centering}p{2cm} |>{\centering\arraybackslash}p{2cm}|}
\hline\hline
$J^{PC}$ & $M_{X_b}$ $(\rm{GeV})$ & $J^{PC}$ & $M_{X_b}$ $(\rm{GeV})$ \\ \hline
$j^{0^{+-}}$ & $19.40^{+0.17}_{-0.17}$ & $j_{A}^{1^{-+}}$ & $19.42^{+0.15}_{-0.16}$ \\ \hline
$j_{A}^{1^{--}}$ & $19.42^{+0.15}_{-0.16}$      & $j_{C}^{1^{-+}}$ & $19.51^{+0.16}_{-0.16}$  \\ \hline
$j_{C}^{1^{--}}$ & $19.52^{+0.16}_{-0.16}$      & $j_{D}^{1^{-+}}$ & $19.34^{+0.15}_{-0.16}$  \\ \hline
$j_{B}^{1^{+-}}$ & $19.20^{+0.16}_{-0.15}$      & $j_{B}^{1^{++}}$ & $19.21^{+0.15}_{-0.15}$  \\ \hline
$j_{D}^{1^{+-}}$ & $19.36^{+0.17}_{-0.16}$      & \textbf{---}     &  \textbf{---}     \\
\hline\hline
\end{tabular}
\end{center}
\caption{Theoretical predictions for hadronic masses of tetrabottom hybrid states across various $J^{PC}$ quantum numbers.}
\label{tab3}
\end{table}

\section{Decay analysis}

\begin{figure}[htb]
    \centering
    \includegraphics[width=1.0\linewidth]{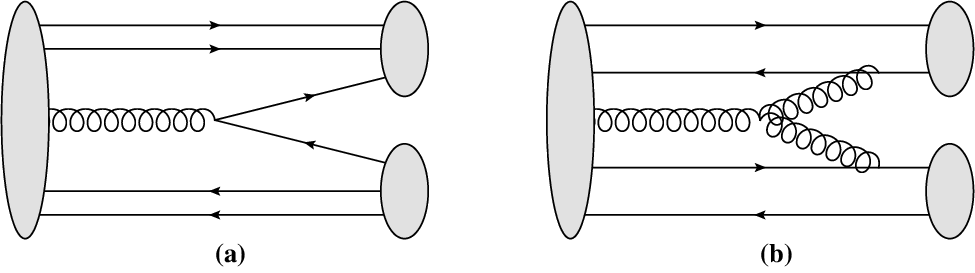}
    \caption{Schematic diagram of two possible decay mechanisms for fully heavy tetraquark hybrid states.}
    \label{decay mechanisms}
\end{figure}

The analysis of decay mechanisms in fully charmed tetraquark hybrid states reveals distinctive behavior in their strong decays. Two primary decay pathways are possible: either the dynamical gluon within the hybrid splits into a quark–antiquark pair, resulting in a pair of doubly charmed baryons (Fig.~\ref{decay mechanisms}(a)); or it splits into two gluons that are absorbed by the quark and antiquark in the hybrid, ultimately yielding two charmonium states (Fig.~\ref{decay mechanisms}(b)).

Based on these mechanisms, we systematically summarize the typical decay channels of tetracharm hybrid states in Table \ref{tab4}. As shown, states with different quantum numbers \(J^{PC}\) predominantly decay via S- and P-wave modes into final states composed of either a pair of doubly charmed baryons or two charmonium mesons. Notably, certain exotic quantum numbers exhibit unique decay patterns: for example, the \(0^{--}\) state can decay via S-wave into \(J/\psi\chi_{c1}\), while its P-wave decays include multiple channels such as \(\eta_c J/\psi\) and \(\chi_{cJ}h_c\).

We emphasize that any new resonance observed in the P-wave double \(J/\psi\) channel must carry the exotic quantum numbers \(1^{-+}\), making it a compelling candidate for the predicted \(1^{-+}\) tetracharm hybrid state. Additionally, states with conventional quantum numbers \(1^{--}\), \(1^{+-}\), and \(1^{++}\) are found to decay not only into two charmonium states but also into doubly charmed baryon pairs such as \(\Xi_{cc}^{++}\bar{\Xi}_{cc}^{--}\) and \(\Omega_{cc}^{+}\bar{\Omega}_{cc}^{-}\).

This systematic analysis of decay patterns offers essential theoretical guidance and observable signatures for identifying and confirming such tetracharm hybrid states in experiments at facilities such as LHCb and CMS.

\begin{table}[h]
\centering
\caption{Possible S-wave and P-wave decay modes of the tetracharm hybird states.}
\begin{tabular}{c|c|c}
\hline
\hline
\(J^{PC}\) & S-wave & P-wave \\
\hline
\(0^{--}\) & \(J/\psi \chi_{c1}\) & $\eta_c J/\psi,\chi_{c1}h_{c}, \chi_{c2}h_{c}, \chi_{c0}h_{c}$ \\
\hline
\(0^{+-}\) & $\chi_{c1}h_{c}$ & $\eta_{c}h_{c}, J/\psi \chi_{c1}, J/\psi \chi_{c2}, J/\psi \chi_{c0}$ \\
\hline
\(1^{--}\) & $\raisebox{-0.4ex}{$\Xi_{cc}^{++}\bar{\Xi}_{cc}^{--}, \Xi_{cc}^{+}\bar{\Xi}_{cc}^{-}, \Omega_{cc}^{+}\bar{\Omega}_{cc}^{-}$}$ & $J/\psi \eta_{c}, \chi_{c0}h_{c}, \chi_{c1}h_{c}, \chi_{c2}h_{c}$\\
           &$J/\psi \chi_{c0}, J/\psi \chi_{c1}, J/\psi \chi_{c2}, \eta_{c}h_{c}$ &\\
\hline
\(1^{+-}\) & $J/\psi \eta_{c}, \chi_{c0}h_{c},\chi_{c1}h_{c},\chi_{c2}h_{c}$ &$\raisebox{-0.4ex}{$\Xi_{cc}^{++}\bar{\Xi}_{cc}^{--}, \Xi_{cc}^{+}\bar{\Xi}_{cc}^{-}, \Omega_{cc}^{+}\bar{\Omega}_{cc}^{-}$}$ \\
& &$J/\psi \chi_{c0}, J/\psi \chi_{c1}, J/\psi \chi_{c2}, \eta_{c}h_{c}$ \\
\hline
\(1^{-+}\) & $J/\psi h_{c}, \eta_{c}\chi_{c1}$ & di-$J/\psi$, di-$\chi_{c1}$, di-$\chi_{c2}$, di-$h_{c}$,  \\
& & $\chi_{c0}\chi_{c1}, \chi_{c1}\chi_{c2}, \chi_{c0}\chi_{c2}$
\\
\hline
\(1^{++}\) & $\chi_{c0}\chi_{c1}, \chi_{c1}\chi_{c2}$ &$\raisebox{-0.4ex}{$\Xi_{cc}^{++}\bar{\Xi}_{cc}^{--}, \Xi_{cc}^{+}\bar{\Xi}_{cc}^{-}, \Omega_{cc}^{+}\bar{\Omega}_{cc}^{-}$}$ \\
& &$J/\psi h_{c}, \eta_{c}\chi_{c0}, \eta_{c}\chi_{c1}, \eta_{c}\chi_{c2}$ \\
\hline
\hline
\end{tabular}
\label{tab4}
\end{table}

Following the same analysis applied to tetracharm hybrids, we present the corresponding decay channels for tetrabottom hybrids in Table \ref{tab5}. Recent experimental progress has been made in the study of fully bottom tetraquark states by the LHCb and CMS collaborations. LHCb conducted a search for $bb\bar{b}\bar{b}$ tetraquark states decaying into $\Upsilon(1S)\mu^{+}\mu^{-}$—with a final state signature of four muons ($\mu^{+}\mu^{-}\mu^{+}\mu^{-}$)—using proton-proton collision data at center-of-mass energies of $\sqrt{s}$ = 7, 8, and 13 TeV, and set upper limits on the production cross section of such states~\cite{LHCb:2018uwm}. In a complementary search, a preliminary analysis of CMS data, following their 2017 reference measurement of $\Upsilon(1S)$-pair production in $pp$ collisions at $\sqrt{s} = 8$ TeV~\cite{CMS:2016liw}, identified an anomalous signal at 18.4 GeV in the $\Upsilon(1S)\ell^+\ell^-$ decay channel with a global significance of 3.6$\sigma$~\cite{Yi:2018fxo}. The future confirmation of this anomaly would provide a strong indication for a fully bottom tetraquark state ($bb\bar{b}\bar{b}$).

We note that the masses of the tetrabottom hybrids predicted in this work exceed the experimental value of 18.4 GeV. This discrepancy suggests that the observed signal around 18.4 GeV likely originates from a conventional four-bottom tetraquark state without gluonic excitation. In contrast, our predicted mass range for the tetrabottom hybrid state with an additional constituent gluon is 19.2-19.5 GeV. The mass increase of approximately 1 GeV due to the constituent gluon is consistent with simple qualitative estimates. Therefore, we strongly encourage further experimental investigations in this higher-energy region to verify the existence of these predicted hybrid states.

\begin{table}[h]
\centering
\caption{Possible S-wave and P-wave decay modes of the tetrabottom hybrid states.}
\begin{tabular}{c|c|c}
\hline
\hline
\(J^{PC}\) & S-wave & P-wave \\
\hline
\(0^{--}\) & $\Upsilon(1S)\chi_{b1}$ & $\eta_{b}\Upsilon(1S), \chi_{b0}h_{b}, \chi_{b1}h_{b}, \chi_{b2}h_{2}$\\
\hline
\(0^{+-}\) & $\chi_{b1}h_{b}$ & $\eta_{b}h_{b}, \Upsilon(1S)\chi_{b0},\Upsilon(1S)\chi_{b1}, \Upsilon(1S)\chi_{b2}$\\
\hline
\(1^{--}\) & $\raisebox{-0.4ex}{$\Xi_{bb}^{0}\bar{\Xi}_{bb}^{0}, \Xi_{bb}^{-}\bar{\Xi}_{bb}^{+}, \Omega_{bb}^{-}\bar{\Omega}_{bb}^{+}$}$ & $\eta_{b}\Upsilon(1S), h_{b}\chi_{c0}, h_{b}\chi_{c1}, h_{b}\chi_{c2}$\\
           &$\eta_{b}h_{b}, \Upsilon(1S)\chi_{b0}, \Upsilon(1S)\chi_{b1}, \Upsilon(1S)\chi_{b2}$&\\
\hline
\(1^{+-}\) & $\eta_{b}\Upsilon(1S), h_{b}\chi_{b0}, h_{b}\chi_{b1}, h_{b}\chi_{b2}$ &$\raisebox{-0.4ex}{$\Xi_{bb}^{0}\bar{\Xi}_{bb}^{0}, \Xi_{bb}^{-}\bar{\Xi}_{bb}^{+}, \Omega_{bb}^{-}\bar{\Omega}_{bb}^{+}$}$ \\
& & $\eta_{b}h_{b}, \Upsilon(1S)\chi_{b0}, \Upsilon(1S)\chi_{b1}, \Upsilon(1S)\chi_{b2}$\\
\hline
\(1^{-+}\) & $\eta_{b}\chi_{b1}, \Upsilon(1S)h_{b}$ & di-$\Upsilon(1S)$, di-$\chi_{b1}$, di-$\chi_{b2}$, di-$h_{b}$\\
& & $\chi_{b0}\chi_{b1}, \chi_{b0}\chi_{b2}, \chi_{b1}\chi_{b2}$
\\
\hline
\(1^{++}\) & $\chi_{b0}\chi_{b1},\chi_{b1}\chi_{b2}$ & $\raisebox{-0.4ex}{$\Xi_{bb}^{0}\bar{\Xi}_{bb}^{0}, \Xi_{bb}^{-}\bar{\Xi}_{bb}^{+}, \Omega_{bb}^{-}\bar{\Omega}_{bb}^{+}$}$\\
&  &$\eta_{b}\chi_{b0}, \eta_{b}\chi_{b1}, \eta_{b}\chi_{b2}, \Upsilon(1S)h_{b}$ \\
\hline
\hline
\end{tabular}
\label{tab5}
\end{table}

\section{Conclusions}
This study employs QCD sum rules to methodically investigate the mass spectrum of tetraquark hybrid states within the $8_{[c\bar{c}]}\otimes 8_{[G]}\otimes 8_{[c\bar{c}]}$ color configuration. Building upon our previous calculations of $0^{++}$ and $0^{-+}$ states, we have formulated a complete set of 18 interpolating currents encompassing all possible quantum number states with spins 0 and 1. By implementing operator product expansion techniques while incorporating nonperturbative contributions up to dimension six, we derive systematic mass predictions for tetracharm hybrid states: states with quantum numbers $0^{+-}$, $1^{--}$, and $1^{-+}$ concentrate within the $7.2-7.3$ GeV range, whereas the $1^{+-}$ and $1^{++}$ states are distributed across $6.9-7.1$ GeV. These numerical outcomes, comprehensively tabulated in Table \ref{tab2}, furnish novel theoretical perspectives for interpreting the di-$J/\psi$ structure observed in LHCb experiments.

Moreover, we extend our investigation to the bottom-quark sector, predicting the mass spectrum of tetrabottom hybrid states with $0^{+-}$, $1^{--}$, and $1^{-+}$ states positioned at $19.4-19.5$ GeV, and $1^{+-}$ and $1^{++}$ states located at $19.2-19.3$ GeV. These results, systematically compiled in Table \ref{tab3}, necessitate experimental corroboration in forthcoming studies.

%%%%%%%%%%%%%%%%%%%%%%%%%%%%%%%%%%%%%%%%%%%%%%%%%%%%%%%%%%%%%%%%%%%%%%
\vspace{.7cm} {\bf Acknowledgments} \vspace{.3cm}

This work is supported by the National Natural Science Foundation of China (NSFC) under Grant No. 11975090, 12475087 and 12235008; the Natural Science Foundation of Hebei Province under Grant No. A2023205038; the S$\&$T Program of Hebei (Grant No. 22567617H) and the Hebei Normal University Graduate Student Innovation Funding Project (Project No. XCXZZBS202503).

%%%%%%%%%%%%%%%%%%%%%%%%%%%%%%%%%%%%%%%%%%%%%%%%%%%%%%%%%%%%%%%%%%%%%%%

\begin{widetext}

  \newpage

\appendix \label{appendix}

\textbf{Appendix}

In this appendix, the spectral densities $\rho^{\text{OPE}}(s)$ corresponding to the currents $j^{0^{+-}}$, $j_{A}^{1^{--}}$, $j^{1^{+-}}_{B}$, $j^{1^{++}}_{B}$ and $j^{1^{-+}}_{D}$ are explicitly given as follows:
\begin{eqnarray}
\rho_{0^{+-}}^{\text{pert}}(s) &=& \frac{g_{s}^{2}}{2^{13}\times  5\pi^{8}}\int_{x_{i}}^{x_{f}} dx\int_{y_{i}}^{y_{f}} dy\int_{z_{i}}^{z_{f}}dz \left\{ F_{xyz}^{6}(1-x-y-z)xyz\right\} \nonumber\\
&-&\frac{3g_{s}^{2}}{2^{13}\times 5\pi^{8}}\int_{x_{i}}^{x_{f}} dx\int_{y_{i}}^{y_{f}} dy\int_{z_{i}}^{z_{f}}dz \int_{\omega_{i}}^{\omega_{f}}d\omega\left\{ F_{xyz\omega}^{5}m_{Q}^{2}(xy-z\omega)\right\},
\end{eqnarray}

\begin{eqnarray}
\rho_{0^{+-}}^{\langle GG\rangle}(s) &=& \frac{\langle g_{s}^{2}GG\rangle}{2^{12}\times\pi^{6}} \int_{x_{i}}^{x_{f}} dx\int_{y_{i}}^{y_{f}} dy \int_{z_{i}}^{z_{f}} dz\left\{m_{Q}^{2}(6m_{Q}^{2}+(2F_{xyz}-3s)(x(y+z)\right.\nonumber\\
&+&\left.z(y+z-1)))\right\},
\end{eqnarray}

\begin{eqnarray}
\rho_{0^{+-}}^{\langle GGG \rangle }(s) &=& \frac{\langle g_{s}^{3}GGG\rangle }{2^{13}\times\pi^{6}} \int_{x_{i}}^{x_{f}} dx\int_{y_{i}}^{y_{f}} dy\int_{z_{i}}^{z_{f}} dz \frac{1}{xyz(x+y+z-1)}\left\{ 3F_{xyz}m_{Q}^{2}(x^{2}(y+z)\right.\nonumber\\
&+&\left.x(y^{2}+y(2z-1)+(z-1)z)+yz(y+z-1))(2m_{Q}^{2}+ (F_{xyz}-s)(x(y+z)\right.\nonumber\\
&+&\left.z(y+z-1)))\right\}-3(-2F_{xyz}mQ^{4}+mQ^{4}s+4F_{xyz}^{3}xyz(x+y+z-1) \nonumber\\
&-& 18F_{xyz}^{2}sxyz(x+y+z-1)+12F_{xyz}s^{2}xyz(x+y+z-1)\nonumber\\
&-& s^{3}xyz(x+y+z-1))
\end{eqnarray}

\begin{eqnarray}
\rho_{1^{--},\,A}^{\text{pert}}(s) &=& \frac{g_{s}^{2}}{2^{12}\times  15\pi^{8}}\int_{x_{i}}^{x_{f}} dx\int_{y_{i}}^{y_{f}} dy\int_{z_{i}}^{z_{f}}dz \left\{ -F_{xyz}^{6}(x+y+z-1)xyz\right\} \nonumber\\
&+&\frac{g_{s}^{2}}{2^{12}\times 15\pi^{8}}\int_{x_{i}}^{x_{f}} dx\int_{y_{i}}^{y_{f}} dy\int_{z_{i}}^{z_{f}}dz \int_{\omega_{i}}^{\omega_{f}}d\omega\left\{ F_{xyz\omega}^{3}(F_{xyz\omega}^{2}(F_{xyz\omega}\right.\nonumber\\
&-&\left.12s)xyz\omega - F_{xyz\omega}mQ^{2}(x-y)(z-\omega)(F_{xyz\omega}(-5+x+y+z+\omega) \right.\nonumber\\
&-&\left.10s(-1+x+y+z +\omega))+10mQ^{2}(-1+x+y+z+\omega)(8s(-1\right.\nonumber\\
&+&\left.x+y+z+\omega)-F_{xyz\omega} (x+y+z+\omega)) - 2F_{xyz\omega}mQ^{2}(F_{xyz\omega}(3x^{2}y\right.\nonumber\\
&-&\left.z\omega(-1+2y+2z+2\omega)+ x(3y^{2}-2z\omega+3y(-2+z+\omega)))\right.\nonumber\\
&-&\left.10s(3x^{2}y-2z\omega(-1+y+z+\omega) x(3y^{2}-2z\omega\right.\nonumber\\
&+&\left. 3y(-1+ z+\omega)))))\right\},
\end{eqnarray}

\begin{eqnarray}
\rho_{1^{--},\,A}^{\langle GG\rangle}(s) &=& \frac{\langle g_{s}^{2}GG\rangle}{2^{15}\times3\pi^{6}} \int_{x_{i}}^{x_{f}} dx\int_{y_{i}}^{y_{f}} dy \int_{z_{i}}^{z_{f}} dz\left\{F_{xyz}^{2}(-36mQ^{4} - (9F_{xyz}^{2}\right.\nonumber\\
&-& \left.28F_{xyz}s+12s^{2})xyz(-1+x+y+z)-6mQ^{2}(F_{xyz}-2s)\right.\nonumber\\
&&\left.(x-y)(-1+x+y +2z) + 12mQ^{2}(2F_{xyz}z(-1+y+z)\right.\nonumber\\
&-&\left.3sz(-1+y+z)+F_{xyz}(y+2z)-sx(y+3z)))\right\},
\end{eqnarray}

\begin{eqnarray}
\rho_{1^{--},\,A}^{\langle GGG \rangle }(s) &=& \frac{\langle g_{s}^{3}GGG\rangle }{2^{16}\times\pi^{6}} \int_{x_{i}}^{x_{f}} dx\int_{y_{i}}^{y_{f}} dy\int_{z_{i}}^{z_{f}} dz  -\frac{1}{z(x+y+z-1)}\left\{mQ^{2}(3F_{xyz}\right.\nonumber\\
&-&\left.4s)(x-y)(-1+x+y+2z)(yz(-1+y+z) + x^{2}(y+z)\right.\nonumber\\
&+&\left.x(y^{2}+(-1+z)z+y(-1+2z)))-2(x+y)(6mQ^{4}+(3F_{xyz}^{2} \right.\nonumber\\
&-&\left.7F_{xyz}s+2s^{2}) xyz(-1 +x+y+z)+mQ^{2}(-6F_{xyz}z(-1+y+z)\right.\nonumber\\
&+&\left.6sz(-1+y+z)-3F_{xyz}x(y+2z) +2sx(y+3z)))
\right\} \nonumber\\
&+& 4\left\{ -12F_{xyz}^{3}xyz(-1+x+y+z)+F_{xyz}^{2} (64sxyz(-1+x+y+z)\right.\nonumber\\
&+&\left.9mQ^{2}(z(-1+y+z) + x(y+z)))+4s(mQ^{4}+s^{2}xyz(-1+x\right.\nonumber\\
&+&\left.y+z)+mQ^{2}s(z(-1+y+z)+x(y+z)))-2F_{xyz} (3mQ^{4}\right.\nonumber\\
&+&\left.23s^{2}xyz(-1+x+y+z)+11mQ^{2}s(z(-1+y+z)\right.\nonumber\\
&+&\left.x(y+z)))\right\},
\end{eqnarray}

\begin{eqnarray}
\rho_{1^{+-},\,B}^{\text{pert}}(s) &=& \frac{g_{s}^{2}}{2^{15}\times  15\pi^{8}}\int_{x_{i}}^{x_{f}} dx\int_{y_{i}}^{y_{f}} dy\int_{z_{i}}^{z_{f}}dz \left\{ -F_{xyz}^{6}(x+y+z-1)xyz\right\} \nonumber\\
&+&\frac{g_{s}^{2}}{2^{15}\times 15\pi^{8}}\int_{x_{i}}^{x_{f}} dx\int_{y_{i}}^{y_{f}} dy\int_{z_{i}}^{z_{f}}dz \int_{\omega_{i}}^{\omega_{f}}d\omega\left\{ F_{xyz\omega}^{3}(F_{xyz\omega}^{2}(F_{xyz\omega}\right.\nonumber\\
&-&\left.12s)xyz\omega -10mQ^{4}(-1+x+y+z+\omega)(8s(-1+x+y+z+\omega)\right.\nonumber\\
&-&\left. F_{xyz\omega}(x+y+z+\omega))+2F_{xyz\omega}mQ^{2}(F_{xyz\omega}(3x^{2}y + z\omega(-1+2y\right.\nonumber\\
&+&\left.2z+2\omega) + x(3y^{2}+2z\omega+3y(-2+z+\omega)))-10s(3x^{2}y\right.\nonumber\\
&+&\left.2z\omega(-1+y+z+\omega) + x(3y^{2}+2z\omega+3y(-1+z+\omega))))\right\},
\end{eqnarray}

\begin{eqnarray}
\rho_{1^{+-},\,B}^{\langle GG\rangle}(s) &=& \frac{\langle g_{s}^{2}GG\rangle}{2^{15}\times3\pi^{6}} \int_{x_{i}}^{x_{f}} dx\int_{y_{i}}^{y_{f}} dy \int_{z_{i}}^{z_{f}} dz\left\{-F_{xyz}^{2}(-36mQ^{4} + (9F_{xyz}^{2}\right.\nonumber\\
&-& \left.28F_{xyz}s+12s^{2})xyz(-1+x+y+z)+12mQ^{2}(-sx(y-3z)\right.\nonumber\\
&+&\left.F_{xyz}x(y-2z)- 2F_{xyz}z(-1+y+z)+3sz(-1+y+z))\right\},
\end{eqnarray}

\begin{eqnarray}
\rho_{1^{+-},\,B}^{\langle GGG \rangle }(s) &=& \frac{\langle g_{s}^{3}GGG\rangle }{2^{15}\times\pi^{6}} \int_{x_{i}}^{x_{f}} dx\int_{y_{i}}^{y_{f}} dy\int_{z_{i}}^{z_{f}} dz  -\frac{1}{xy}\left\{F_{xyz}(x+y)(-6mQ^{4}\right.\nonumber\\
&+&\left.3F_{xyz}^{2}-7F_{xyz}s + 2s^{2})xyz(-1+x+y+z)+mQ^{2}(-2sx(y\right.\nonumber\\
&-&\left.3z) + 3F_{xyz}x(y-2z) - 6F_{xyz}z(-1+y+z) + 6sz(-1+y+z))\right\} \nonumber\\
&+& 2\left\{-12F_{xyz}^{3}xyz(-1+x+y+z)+F_{xyz}^{2} (64sxyz(-1+x+y+z)\right.\nonumber\\
&-&\left. 9mQ^{2}(x(y-z)-z(-1+y+z)))+4s(-mQ^{4}+s^{2}xyz(-1\right.\nonumber\\
&+&\left.x+y+z) + 22mQ^{2}s(x(y-z)-z(-1+y+z))) + 4s(-mQ^{4}\right.\nonumber\\
&+&\left.s^{2}xyz(-1+x+y+z)+mQ^{2}s (x(-y+z)+z(-1+y+z)))\right\},
\end{eqnarray}

\begin{eqnarray}
\rho_{1^{++},\,B}^{\text{pert}}(s) &=& \frac{g_{s}^{2}}{2^{15}\times  3^{3}\pi^{8}}\int_{x_{i}}^{x_{f}} dx\int_{y_{i}}^{y_{f}} dy\int_{z_{i}}^{z_{f}}dz \left\{ -F_{xyz}^{6}(x+y+z-1)xyz\right\} \nonumber\\
&+&\frac{g_{s}^{2}}{2^{15}\times 3^{3}\pi^{8}}\int_{x_{i}}^{x_{f}} dx\int_{y_{i}}^{y_{f}} dy\int_{z_{i}}^{z_{f}}dz \int_{\omega_{i}}^{\omega_{f}}d\omega\left\{ F_{xyz\omega}^{3}(F_{xyz\omega}^{2}(F_{xyz\omega}\right.\nonumber\\
&-&\left.12s)xyz\omega -10mQ^{4}(-1+x+y+z+\omega)(8s(-1+x+y+z+\omega)\right.\nonumber\\
&-&\left. F_{xyz\omega}(x+y+z+\omega))+2F_{xyz\omega}mQ^{2}(F_{xyz\omega}(3x^{2}y + z\omega(-1+2y\right.\nonumber\\
&+&\left.2z+2\omega) + x(3y^{2}+2z\omega+3y(-2+z+\omega)))-10s(3x^{2}y\right.\nonumber\\
&+&\left.2z\omega(-1+y+z+\omega) + x(3y^{2}+2z\omega+3y(-1+z+\omega))))\right\},
\end{eqnarray}

\begin{eqnarray}
\rho_{1^{++},\,B}^{\langle GG\rangle}(s) &=& \frac{\langle g_{s}^{2}GG\rangle}{2^{15}\times3\pi^{6}} \int_{x_{i}}^{x_{f}} dx\int_{y_{i}}^{y_{f}} dy \int_{z_{i}}^{z_{f}} dz\left\{-F_{xyz}^{2}(-36mQ^{4} + (9F_{xyz}^{2}\right.\nonumber\\
&-& \left.28F_{xyz}s+12s^{2})xyz(-1+x+y+z)+12mQ^{2}(-sx(y-3z)\right.\nonumber\\
&+&\left.F_{xyz}x(y-2z)- 2F_{xyz}z(-1+y+z)+3sz(-1+y+z))\right\},
\end{eqnarray}

\begin{eqnarray}
\rho_{1^{++},\,B}^{\langle GGG \rangle }(s) &=& \frac{\langle g_{s}^{3}GGG\rangle }{2^{15}\times3^{2}\pi^{6}} \int_{x_{i}}^{x_{f}} dx\int_{y_{i}}^{y_{f}} dy\int_{z_{i}}^{z_{f}} dz  -\frac{1}{xy}\left\{F_{xyz}(x+y)(-6mQ^{4}\right.\nonumber\\
&+&\left.3F_{xyz}^{2}-7F_{xyz}s + 2s^{2})xyz(-1+x+y+z)+mQ^{2}(-2sx(y\right.\nonumber\\
&-&\left.3z) + 3F_{xyz}x(y-2z) - 6F_{xyz}z(-1+y+z) + 6sz(-1+y+z))\right\} \nonumber\\
&-& 2\left\{12F_{xyz}^{3}xyz(-1+x+y+z)+F_{xyz}^{2} (-64sxyz(-1+x+y+z)\right.\nonumber\\
&+&\left. 9mQ^{2}(x(y-z)-z(-1+y+z)))-4s(-mQ^{4}+s^{2}xyz(-1\right.\nonumber\\
&+&\left.x+y+z) + mQ^{2}s(x(-y+z)+z(-1+y+z))) + F_{xyz}(-6mQ^{4}\right.\nonumber\\ &+&\left.46s^{2}xyz(-1+x+y+z)-22mQ^{2}s(x(y-z)\right.\nonumber\\
&-&\left.z(-1+y+z)))\right\},
\end{eqnarray}

\begin{eqnarray}
\rho_{1^{-+},\,D}^{\text{pert}}(s) &=& \frac{g_{s}^{2}}{2^{15}\times  15\pi^{8}}\int_{x_{i}}^{x_{f}} dx\int_{y_{i}}^{y_{f}} dy\int_{z_{i}}^{z_{f}}dz \left\{ -F_{xyz}^{6}(x+y+z-1)xyz\right\} \nonumber\\
&+&\frac{g_{s}^{2}}{2^{15}\times 15\pi^{8}}\int_{x_{i}}^{x_{f}} dx\int_{y_{i}}^{y_{f}} dy\int_{z_{i}}^{z_{f}}dz \int_{\omega_{i}}^{\omega_{f}}d\omega\left\{ F_{xyz\omega}^{3}(F_{xyz\omega}^{2}(F_{xyz\omega}\right.\nonumber\\
&-&\left.12s)xyz\omega-F_{xyz\omega}mQ^{2}(x+y)(z+\omega)(F_{xyz\omega} (-5+x+y+z+\omega)\right.\nonumber\\
&-&\left.10s(-1+x+y+z+\omega))+10mQ^{4}(-1+x+y+z+\omega)\right.\nonumber\\
&\times&\left.(8s (-1+x+y+z+\omega)-F_{xyz\omega}(x+y+z+\omega))\right.\nonumber\\ &+&\left. 2F_{xyz\omega}mQ^{2}(F_{xyz\omega}(3x^{2}y-z\omega (-1+2y+2z+2\omega)\right.\nonumber\\
&+&\left. x(3y^{2}-2z\omega+3y(-2+z+\omega))) + 10s(3x^{2}y-2z\omega(-1+y\right.\nonumber\\
&+&\left.z+\omega)+x(3y^{2}-2z\omega+3y(-1+z+\omega)))))\right\},
\end{eqnarray}

\begin{eqnarray}
\rho_{1^{-+},\,D}^{\langle GG\rangle}(s) &=& \frac{\langle g_{s}^{2}GG\rangle}{2^{15}\times3\pi^{6}} \int_{x_{i}}^{x_{f}} dx\int_{y_{i}}^{y_{f}} dy \int_{z_{i}}^{z_{f}} dz\left\{F_{xyz}^{2}(-36mQ^{4}+6mQ^{2}(F_{xyz}\right.\nonumber\\
&-&\left.2s)(-1+x+y)(x+y) - (9F_{xyz}^{2} -28F_{xyz}s +12s^{2})xyz(-1\right.\nonumber\\
&+&\left.x+y+z)-12mQ^{2}(2F_{xyz}z (-1+y+z)-3sz(-1+y+z)\right.\nonumber\\
&+&\left.F_{xyz}x(y+2z)-sx(y+3z)))\right\},
\end{eqnarray}

\begin{eqnarray}
\rho_{1^{-+},\,D}^{\langle GGG \rangle }(s) &=& \frac{\langle g_{s}^{3}GGG\rangle }{2^{16}\times\pi^{6}} \int_{x_{i}}^{x_{f}} dx\int_{y_{i}}^{y_{f}} dy\int_{z_{i}}^{z_{f}} dz  \left\{\frac{F_{xyz}}{xy}(-2(x+y)(6mQ^{4}\right.\nonumber\\
&+&\left.(3F_{xyz}^{2}-7F_{xyz}s + 2s^{2})xyz(-1+x+y+z)+mQ^{2}(6F_{xyz}z\right.\nonumber\\
&\times&\left.(-1+y+z) - 6sz(-1+y+z)+3F_{xyz}x(y+2z)-2sx(y\right.\nonumber\\
&+&\left.3z))) + \frac{1}{z(x+y+z-1)} mQ^{2}(3F_{xyz}-4s)((-1+y)y^{2}z\right.\nonumber\\
&\times&\left.(-1+y+z)+x^{4}(y+z)+x^{3}(3y^{2} + (-2+z)z+y(-2\right.\nonumber\\
&+&\left.4z))+xy^{2}(1+y^{2}-4z+3z^{2}+y(-2+4z)) + x^{2}(3y^{3}\right.\nonumber\\
&-&\left.(-1+z)z+y^{2}(-4+6z) + y(1-4z+3z^{2})))\right\}\nonumber\\
&-& 4\left\{ (12F_{xyz}^{3}xyz(-1+x+y+z)+F_{xyz}^{2}(-64sxyz(-1\right.\nonumber\\
&+&\left.x+y+z) + 9mQ^{2}(z(-1+y+z)+x(y+z)))\right.\nonumber\\
&+&\left. F_{xyz}(6mQ^{4} + 46s^{2}xyz(-1+x+y+z) - 22mQ^{2}s\right.\nonumber\\
&\times&\left.(z(-1+y+z) + x(y+z))) - 4s(mQ^{4}+s^{2}xyz(-1\right.\nonumber\\
&+&\left.x+y+z) -mQ^{2}s(z(-1+y+z)+x(y+z))))\right\},
\end{eqnarray}

where we have used the following definitions:
\begin{eqnarray}
  F_{xyz} &=& m_{Q}^{2}f_{xyz} - s, \\
  F_{xyz\omega} &=& \bigg(\frac{1}{x}+\frac{1}{y}+\frac{1}{z}+\frac{1}{w}  \bigg)m_Q^2-s\;,\\
  x_{f/i}&=&\bigg[\bigg( 1-\frac{8m_Q^2}{s} \bigg) \pm \sqrt{\bigg( 1-\frac{8m_Q^2}{s} \bigg)^2-\frac{4m_Q^2}{s}}\bigg] \bigg/2\;,\\
y_{f/i}&=&\bigg[ 1+2 x +\frac{3 s x^2}{m_Q^2-s x} \pm \sqrt{\frac{[m_Q^2+s x (x-1)][(8x+1)m_Q^2+s x(x-1)]}{(m_Q^2-s x)^2}}  \bigg] \bigg/2\;,\\
z_{f/i}&=&\bigg[(1-x-y)\pm \sqrt{\frac{(x+y-1)[m_Q^2(x+y-(x -y)^2)+s x y(x+y-1)]}{s x y-(x+y) m_Q^2}} \bigg]\bigg/2\;,\\
w_{-}&=&\frac{x y z m_Q^2}{s x y z -(x y +y z + x z)m_Q^2}\;,w_{+}=1-x-y-z\;,
\end{eqnarray}
where $\hat{s} = \frac{s}{m_{Q}^{2}}$.
\end{widetext}

\end{document}